\newif\ifANONYMOUS
\ANONYMOUSfalse
\ANONYMOUStrue

\ifANONYMOUS
\documentclass[manuscript]{acmart}
\else
\documentclass[manuscript]{acmart}
\fi

\usepackage{longtable}
\usepackage{float}
\usepackage{subcaption}
\usepackage{comment}
\AtBeginDocument{%
  }

\setcopyright{acmlicensed}
\copyrightyear{2026}
\acmYear{2026}
\acmDOI{XXXXXXX.XXXXXXX}

\usepackage{algpseudocode}
\usepackage[normalem]{ulem} 
\usepackage{xspace}
\usepackage{multirow} 
\usepackage{balance} 
\usepackage{fancyvrb} 
\usepackage{caption}
\usepackage{booktabs} 

\usepackage{enumitem}
\setlist[itemize]{leftmargin=*,noitemsep,topsep=0pt}
\setlist[enumerate]{leftmargin=*}

\usepackage{todonotes}
\usepackage{soul}
\usepackage{xcolor}

\newif\ifDEBUG
\DEBUGtrue 

\ifDEBUG
    \newcommand{\DL}[1]{\todo[color=cyan,inline]{DL: #1}}
    \newcommand{\JD}[1]{\todo[color=yellow,inline]{JD: #1}}
    \newcommand{\BC}[1]{\todo[color=orange,inline]{BC: #1}}
    
\else
    \newcommand{\DL}[1]{}
    \newcommand{\JD}[1]{}
    \newcommand{\BC}[1]{}
    
\fi

\usepackage{hyperref}

\usepackage{cleveref}
\crefformat{section}{\S#2#1#3}
\crefname{figure}{Figure}{Figures}
\crefname{table}{Table}{Tables}
\crefname{theorem}{Theorem}{Theorems}
\crefname{thm}{Theorem}{Theorems}
\crefname{lemma}{Lemma}{Lemmata}
\crefname{equation}{Eqt.}{Eqts.}
\crefformat{Grammar}{Grammar #1}
\crefname{appendix}{Appendix}{Appendices}
\crefname{listing}{Listing}{Listings}

\usepackage{url}

\usepackage{tcolorbox}

\makeatletter
\newcommand{\linebreakand}{%
  \end{@IEEEauthorhalign}
  \hfill\mbox{}\par
  \mbox{}\hfill\begin{@IEEEauthorhalign}
}
\makeatother

\usepackage{pifont}

 \usepackage{pifont}
\begin{document}

\title{Is US Defense Acquisition Ready to Acquire AI-Enabled Capabilities?}
\subtitle{Assessing the DoD Software Acquisition Pathway Through a Scenario-Based Policy Analysis}
\author{Daniel Lugo}
\email{lugod@purdue.edu}
\orcid{0009-0000-2762-6634}
\affiliation{%
  \institution{Purdue University}
  \city{West Lafayette}
  \state{Indiana}
  \country{USA}
}
\author{James C. Davis}
\email{davisjam@purdue.edu}
\orcid{0000-0003-2495-686X}
\affiliation{%
  \institution{Purdue University}
  \city{West Lafayette}
  \state{Indiana}
  \country{USA}
}

\renewcommand{\shortauthors}{Lugo \& Davis}

\begin{abstract}
As AI systems transition from experimental prototypes to mission-critical tools, their dependence on dynamic data, evolving models, and governance raises questions about whether existing acquisition pathways can keep pace. The U.S. Department of Defense has modernized its acquisition processes through the Adaptive Acquisition Framework, with the Software Acquisition Pathway (SWP) serving as the primary mechanism for acquiring software-intensive capabilities. This paper evaluates whether SWP is sufficient to address the unique demands of AI acquisition.

In this work, we perform a scenario-based evaluation that traces a notional AI-enabled program through key SWP planning activities to assess how policy translates into program artifacts and decisions. We use Policy Scenario Analysis to examine whether the SWP-centered governance stack provides sufficient actionable support for AI acquisition. The governance stack provides a viable foundation for iterative delivery and AI testing. However, we identify a recurring actionability problem in the core guidance. AI-specific controls for data provenance, lifecycle management, and human oversight remain distributed across supplemental documents rather than embedded in the program-facing mechanisms through which SWP is executed. This disconnect leaves program offices reliant on inconsistent local interpretation. We conclude by recommending an AI-supporting sub-path and targeted artifact refinements to better bridge this policy-to-artifact gap.

\end{abstract}



\received{ 2026}
\received[revised]{ 2026}
\received[accepted]{ 2026}

\maketitle

\section{Introduction}
\label{sec:intro}

Governments are integrating AI capabilities to improve operational efficiency in domains such as healthcare, education, transportation, finance, and national defense~\cite{education,health,AIgov}. Public-sector delivery is constrained by formal systems of policy, oversight, and budgeting that shape what programs can require, document, approve, and sustain \cite{Kattel2022}. As a result, some governments are developing AI-specific procurement and governance frameworks that make expectations for testing, transparency, risk management, and post-deployment oversight more explicit \cite{japan_procurement_guidelines,eu_ai_act}. For defense acquisition, this raises a broader question about whether existing guidance can accommodate the distinctive demands of AI acquisition.
This tension is particularly acute for AI-enabled capabilities whose performance depends on data quality, model updates, validation practices, and continued monitoring over time~\cite{KarpathySW2,MLTestScore,2023ContinuousAI}. 

In the U.S. Department of Defense (DoD), capability acquisition is organized through the Adaptive Acquisition Framework (AAF)~\cite{DoDI5000_02}. Within AAF, the Software Acquisition Pathway (SWP) was introduced to support rapid, iterative delivery through continuous integration and responsiveness to user needs~\cite{DoDI5000_87}. SWP was designed to move software programs away from rigid, hardware-oriented acquisition approaches and toward processes more consistent with modern software practice. By mandating SWP for software-intensive efforts, DoD has positioned it as the principal organizing framework for software acquisition~\cite{LaPlante2022SWPDBS,DoD2025SoftwareAcquisition}. Acquisition teams must use it within a broader governance environment that includes statutory, regulatory, departmental, and AI-specific guidance~\cite{DAU_AAF_2026,DoDI5000_02}. 

In this study, we consider whether AAF's SWP-centered governance stack provides DoD program offices with sufficiently actionable guidance for the acquisition of AI capabilities. To answer that question, we conduct a scenario-based policy assessment~\cite{cunningham2016_psa} of the DoD acquisition governance stack. The assessment compares a synthetic AI-enabled acquisition scenario with a conventional software acquisition scenario. We decompose both scenarios into discrete acquisition episodes and trace them through the Planning Phase artifacts, decision points, and review structures that organize software acquisition in practice. We then evaluate whether the governing corpus provides explicit, partial, or absent support for AI-relevant acquisition properties at the points where programs must produce artifacts and make decisions.

Our results show that SWP can accommodate the high-level needs of software-intensive AI programs. Its actionability becomes uneven, however, when program success depends on lifecycle controls that are more salient for AI than for conventional software. Across our scenario-based assessment, SWP consistently supports modular execution and general cybersecurity expectations, but offers weaker or absent procedural direction for training-data governance, model traceability, retraining triggers, and performance-drift management. Although supplemental guidance improves coverage in several of these areas, important gaps remain. Overall, our findings indicate that the broader SWP-centered governance stack is unevenly specified for AI-specific acquisition needs.

In summary, our main contributions are:
\begin{itemize}[leftmargin=2em]
    \item \emph{A scenario-based assessment of AI acquisition under SWP.} We apply an established policy-assessment method to the Software Acquisition Pathway and identify where the current governance stack provides explicit, partial, or absent support for AI-specific acquisition demands.
    \item \emph{Recommendations to make SWP more AI-ready.} Where SWP provides partial or absent guidance, we propose targeted enhancements to current guidance, framed as improvements to the existing acquisition-process rather than a wholly separate pathway.
\end{itemize}

\paragraph{Significance:} As AI-enabled capabilities become central to defense missions, acquisition policy must address lifecycle demands that extend beyond conventional software delivery, including data governance, model assurance, ongoing monitoring, and responsible-use oversight. This study contributes both a practical framework for navigating existing guidance when planning AI-enabled acquisitions and an evidence-based basis for refining AAF/SWP guidance where needed.

  

\section{Background and Related Work}
This section provides the background for our policy assessment.~\cref{synthesis_swp} explains the origins and intent of the Adaptive Acquisition Framework (AAF) and the Software Acquisition Pathway (SWP).~\cref{SWvsAI} identifies the key properties that distinguish AI from conventional software and defines the dimensions that we later use in the methodology to assess policy support.~\cref{prior_work} then situates SWP within the broader literature on acquisition reform and AI governance.

\subsection{The Software Acquisition Pathway (SWP)}
\label{synthesis_swp}

\begin{figure}
  \includegraphics[width=0.95\linewidth]{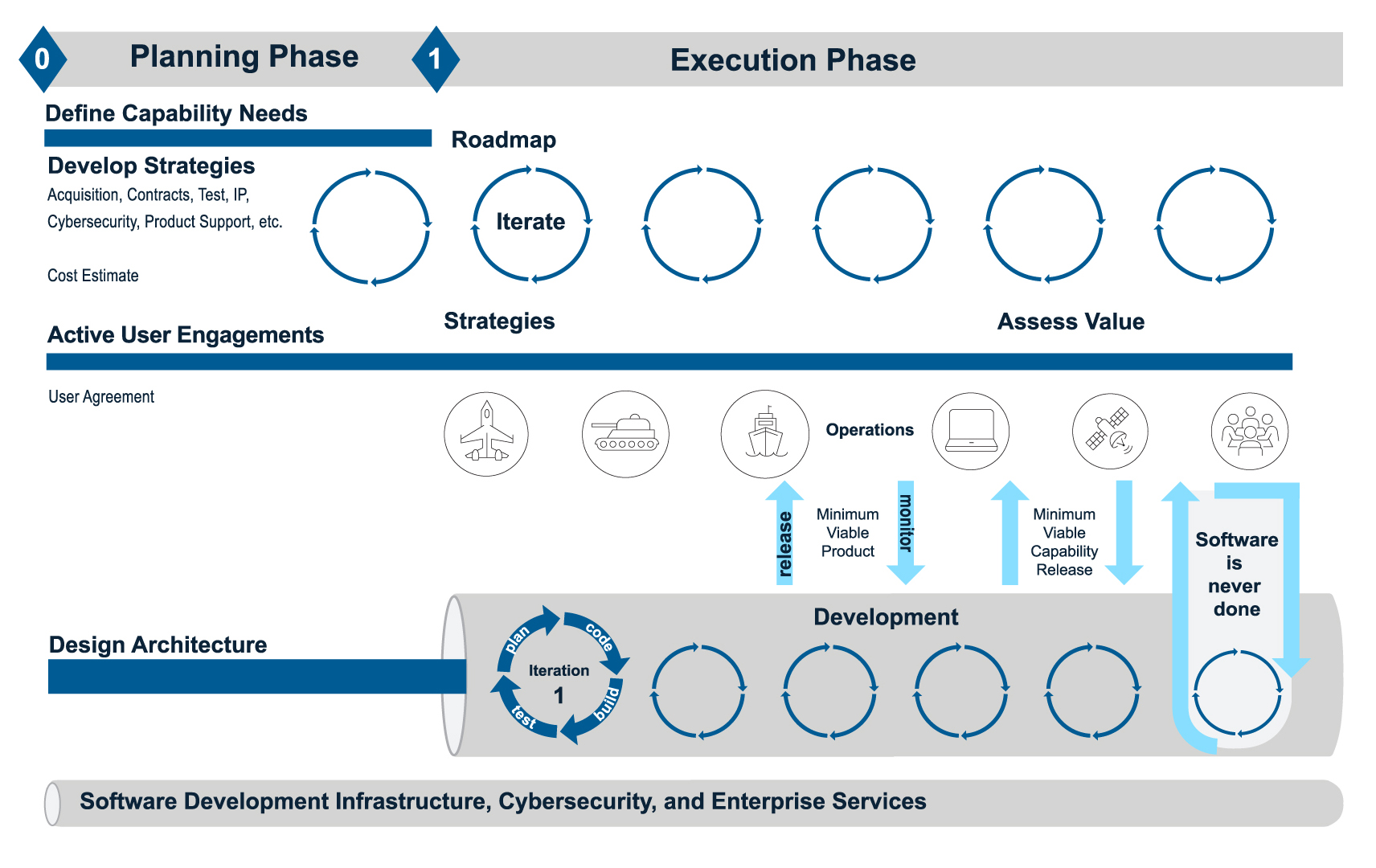}
  \caption{The Software Acquisition Pathway (Reused from DoDI 5000.87 \cite{DoDI5000_87}). This figure illustrates the SWP; it has two main phases (Planning and Execution) and proposes active user engagements and iterative development. There are several plan-code-build-test iterations where software releases can be delivered. }
  \Description{The workflow depicted in this figure highlights the two-stage logic of the SWP. The Planning Phase is characterized by the development of foundational governance artifacts, including the Acquisition, Data, Test, and Cybersecurity Strategies. In the execution stage, the focus shifts to rapid, iterative delivery powered by minimum viable capability releases. }
  \label{fig:SWP}
\end{figure}

The Adaptive Acquisition Framework (AAF) was introduced in 2020 as part of the Department of Defense’s broader acquisition reform effort~\cite{DoDI5000_02}, replacing the long-standing DoDI 5000.02-centered structure that dated back to 2008 and was later revised in 2015~\cite{DoDI5000_02_2008,DoDI5000_02_2015}. DoDI 5000.02 established the policy framework for that change and responded to longstanding criticism that earlier DoD 5000-series processes were too rigid, too slow, and poorly matched to modern software and other emerging technologies~\cite{DoDI5000_02,SWAP2019}. The AAF established six acquisition pathways, including Urgent Capability, Middle Tier, Major Capability Acquisition, and Software Acquisition (SWP), each intended to tailor governance and oversight to the characteristics of different acquisition efforts~\cite{DoDI5000_02}.

AAF did not change acquisition practice through a single top-level policy alone. Its introduction triggered a broader realignment of acquisition guidance across the Department, including pathway-specific instructions and updates to functional and Service-level policies for engineering, testing, cybersecurity, sustainment, and related activities. As a result, acquisition personnel do not execute AAF through a single self-contained instruction. They work instead within a distributed acquisition-governance environment that combines pathway guidance with other authoritative policy instruments.

Within that acquisition environment, DoDI~5000.87 serves as the core instruction for the SWP~\cite{DoDI5000_87}. It organizes the software lifecycle into two primary phases: the Planning Phase and the Execution Phase (see Figure~\ref{fig:SWP}). The Planning Phase is designed to establish the program's technical and governance foundations, requiring the development of a Functional Strategy and a Value Assessment that replace traditional, static requirements with a more dynamic, user-centered approach~\cite{DAU_AAF_Software}. During this phase, acquisition teams must also produce a suite of foundational strategies, including the Acquisition Strategy, Test Strategy, Cybersecurity Strategy, and Product Support Strategy, which define the operational envelope for the capability~\cite{dau_swp_develop_strategies}.

Upon successful completion of these planning milestones, the program enters the Execution Phase, which is characterized by rapid, iterative cycles of development, integration, and delivery~\cite{DAU_AAF_Software}. This phase leverages DevSecOps principles to maintain continuous user engagement and shortened delivery timelines, ideally resulting in a Minimum Viable Product (MVP) and subsequent software increments~\cite{DoDI5000_87}. By institutionalizing Continuous Integration/Continuous Delivery (CI/CD) and continuous Authority to Operate (cATO), the SWP provides the procedural framework intended to keep defense software responsive to changing operational needs~\cite{SWAP2019, DoDI5000_87}.

\begin{table}[t]
    \centering
    \small
    \caption{Documents that acquisition personnel may use when planning and executing software-intensive programs under the Adaptive Acquisition Framework and the Software Acquisition Pathway. Identified via the DAU AAF Acquisition Policies repository \cite{dau_aaf_policies}, the DoD Publications repository \cite{dow_publications_software_pathway}, and the Federal CIO Council policy index \cite{cio_ai_acquisition_policy}. (Snapshot Nov 2025).}
    \label{tab:dod_policies}
    \begin{tabular}{p{2.5cm} p{6.0cm} p{3.8cm} l}
    \toprule
    \textbf{Layer} & \textbf{Title / Identifier} & \textbf{Role in acquisition environment} & \textbf{Date} \\
    \midrule
    \textit{Overarching} 
        & DoDI 5000.02, \emph{Operation of the Adaptive Acquisition Framework} 
        & AAF structure and governance 
        & Jun 2022 \\
    \addlinespace

    \textit{Pathway} 
        & DoDI 5000.87, \emph{Operation of the Software Acquisition Pathway} 
        & SWP procedural anchor 
        & Oct 2020 \\
    \addlinespace

    \textit{Functional} 
        & DoDI 5000.73, \emph{Cost Analysis Guidance and Procedures} 
        & Cost analysis expectations 
        & Oct 2024 \\
        & DoDI 5000.82, \emph{Acquisition of Digital Capabilities} 
        & Digital capability planning 
        & Jun 2023 \\
        & DoDI 5000.83, \emph{Technology and Program Protection to Maintain Technological Advantage} 
        & Program protection expectations 
        & May 2021 \\
        & DoDI 5000.88, \emph{Engineering of Defense Systems} 
        & Systems engineering expectations 
        & Nov 2020 \\
        & DoDI 5000.89, \emph{Test and Evaluation} 
        & T\&E expectations 
        & Nov 2020 \\
        & DoDI 5000.90, \emph{Cybersecurity for Acquisition Decision Authorities and Program Managers} 
        & Cybersecurity expectations 
        & Dec 2020 \\
        & DoDI 5000.91, \emph{Product Support Management for the Adaptive Acquisition Framework} 
        & Sustainment expectations 
        & Nov 2021 \\
        & DoDI 5000.95, \emph{Human Systems Integration in Defense Acquisition} 
        & Human factors 
        & Apr 2022 \\
        & DoDI 5000.97, \emph{Digital Engineering} 
        & Digital engineering practices 
        & Dec 2023 \\
        & DoDI 5010.44, \emph{Intellectual Property (IP) Acquisition and Licensing} & IP and data/software rights strategy & Oct 2019 \\
    \addlinespace

    \textit{Service} 
        & DAFI 63-101/20-101, \emph{Integrated Life Cycle Management} 
        & Air Force implementation overlay 
        & Feb 2024 \\
        & SECNAVINST 5000.2G, \emph{Defense Acquisition System and Joint Capabilities Integration and Development System Implementation} 
        & Navy implementation overlay 
        & Apr 2022 \\
        & Army Regulation 70--1, \emph{Army Acquisition Policy} 
        & Army implementation overlay 
        & Nov 2023 \\
    \addlinespace

    \textit{Supplemental Department} 
        & DoD Responsible AI Strategy and Implementation Pathway 
        & DoD AI governance expectations 
        & 2022 \\
        & CDAO Responsible AI Toolkit 
        & Operational RAI support 
        & 2023 \\
        & CDAO Test and Evaluation Strategy Framework 
        & AI TEVV framing 
        & 2024 \\
        & CDAO AI Test and Evaluation Guidance 
        & AI T\&E implementation detail 
        & 2024 \\
    \addlinespace

    \textit{Supplemental Government-wide} 
        & OMB Memorandum M-25-22, \emph{Driving Efficient Acquisition of Artificial Intelligence in Government} 
        & federal AI acquisition guidance  
        & Apr 2025 \\
    \bottomrule
\end{tabular}
\end{table}

\subsubsection{Documents Used by Acquisition Personnel Under SWP}
\label{subsec:dod_document_environment}
Acquisition personnel do not execute SWP through the pathway instruction alone. They must navigate a broader policy environment that includes the pathway instruction, the functional policies it invokes, and supplemental guidance for engineering, testing, cybersecurity, and sustainment~\cite{DoDI5000_02,dau_aaf_policies,DoDI5000_88,DoDI5000_89,DoDI5000_90}. This is especially relevant for AI-enabled capabilities, which are commonly acquired under SWP when custom software is the primary means of delivering the capability~\cite{DoDI5000_87,DAU_AAF_Software}. Some requirements are addressed through standard software structures, while others rely on cross-cutting mandates from related acquisition functions~\cite{DoDI5000_88,DoDI5000_89,DoDI5000_90}. Although these materials vary in formal procedural status, ranging from mandatory instructions to advisory or explanatory guidance, they collectively constitute the operational guidance environment for a DoD program acquiring software capability~\cite{DoDI5000_02,dau_aaf_policies}.

Table~\ref{tab:dod_policies} summarizes the principal documents that structure the planning and execution of software-intensive programs. Together, they represent the distributed governance stack that defines the compliance and delivery boundaries for acquisition practitioners.

\subsection{AI-Relevant Acquisition Properties}
\label{SWvsAI}

Software now underpins capability delivery and sustainment across the DoD, with a growing share of that investment directed toward AI-enabled capabilities. For FY2025, the DoD requested approximately \$1.8 billion specifically for artificial intelligence initiatives \cite{dod_fy25_ai}; by FY2026, this expanded into a dedicated \$13.4 billion portfolio for autonomy and AI systems \cite{dod_fy26_ai}. To put this in perspective, this dedicated AI and autonomy funding is now equivalent in scale to approximately 20\% of the Department’s entire \$66 billion IT budget. This investment supports a wide range of technologies that extend beyond traditional deterministic software. According to published FY2026 program acquisition cost estimates, these investments are concentrated in several high-priority categories, including aerial autonomy and drones, maritime and underwater systems, and AI-enabled infrastructure and integration~\cite{dod_fy26_weapons}.

AI-enabled capabilities differ from conventional software because they require the continuous interaction of code, data, models, and operations~\cite{sculley_hidden_debt_2015,MLTestScore}. As a result, acquisition success depends not only on delivering executable code, but also on governing data quality, evaluating probabilistic behavior, and maintaining oversight of model evolution. Table~\ref{tab:ai_properties_dimensions} summarizes the AI-relevant properties discussed in this section. These properties were identified through our literature review on how AI-enabled systems differ from conventional software and were selected to highlight distinct acquisition implications for planning, oversight, testing, sustainment, and governance.

\begin{table}[t]
    \centering
    \caption{AI-relevant properties used as evaluation dimensions in the policy assessment, identified through our literature review on how AI-enabled systems differ from conventional software and selected for their acquisition implications.}
    \label{tab:ai_properties_dimensions}
    \begin{small}
    \begin{tabular}{p{4.1cm} p{9.9cm}}
        \toprule
        \textbf{AI-relevant property} & \textbf{Acquisition implication relative to conventional software} \\
        \midrule
        Data as a first-class deliverable &
        Performance and sustainment depend on training, tuning, and evaluation data, making data quality, lineage, and licensing acquisition concerns alongside code and interfaces~\cite{gebru_datasheets_2021,mitchell_model_cards_2019}. \\

        Models evolve and drift &
        Performance may change as missions, environments, or data distributions shift, creating continuing needs for monitoring, retraining, and revalidation~\cite{baylor_tfx_2017,MLTestScore}. \\

        Non-determinism and distribution sensitivity &
        Behavior may vary across inputs and operating conditions, making exhaustive testing infeasible and increasing the importance of robustness evaluation and context-aware assurance~\cite{sculley_hidden_debt_2015,MLTestScore}. \\

        Explainability and accountability &
        Limited interpretability increases the need for traceability, documentation, and provenance linking model behavior to data, design choices, and operational use~\cite{provenance_XAI,scoresheet_XAI}. \\

        Ethics and governance constraints &
        Fairness, privacy, lawful use, and human oversight can determine whether a capability is acceptable for deployment and continued use~\cite{GAO-23-105850,microsoft_2023_ai_procurement}. \\

        Supply-chain and transitive risk &
        Dependence on upstream models, datasets, and toolchains increases provenance, security, and licensing risks across the lifecycle~\cite{intoto2019,slsa_v1_2023,sigstore2021,jiang2023pretrainedreuse}. \\

        Rapid ecosystem churn &
        Fast-moving models, tools, and deployment stacks increase the need for continuous validation, monitoring, and adaptation~\cite{aiindex2025,tfx_mlopsguide_2024,nist_ai_rmf_1_0}. \\

        Compute and specialized hardware &
        Performance, latency, throughput, and cost often depend on accelerator-rich environments, making deployment context more consequential~\cite{jouppi2017tpu,DLAccelSurvey2025}. \\
        \bottomrule
    \end{tabular}
    \end{small}
\end{table}

\begin{itemize}[leftmargin=2em]
    \item \emph{Data as a first-class deliverable.}
    Conventional software often treats data primarily as an input or output. In AI systems, training, tuning, and evaluation data materially shape capability performance, reuse, and sustainment. Data quality, lineage, representativeness, and licensing become acquisition concerns alongside code and interfaces~\cite{gebru_datasheets_2021,mitchell_model_cards_2019}.

    \item \emph{Models evolve and can drift.}
    Conventional software logic usually changes through managed releases. AI models may degrade as environments, missions, or data distributions change, creating ongoing needs for monitoring, retraining, and revalidation~\cite{baylor_tfx_2017,MLTestScore}.

    \item \emph{Assurance is harder under non-determinism and distribution sensitivity.}
    Many AI systems are probabilistic, data-dependent, and difficult to evaluate exhaustively. AI assurance must address robustness, distribution shift, monitoring, and performance across changing operational contexts rather than static test accuracy alone~\cite{sculley_hidden_debt_2015,MLTestScore}.

    \item \emph{Explainability and accountability matter for oversight.}
    Many AI components are only partially interpretable, which complicates assurance, traceability, certification, and human accountability. This increases the importance of documentation and provenance-aware mechanisms that connect model behavior to underlying data, design choices, and operational use~\cite{provenance_XAI,scoresheet_XAI}.

    \item \emph{Ethics and governance become operational requirements.}
    Fairness, bias mitigation, privacy, lawful use, and human oversight are not optional enhancements; they can determine whether an AI capability is acceptable for deployment and continued use~\cite{GAO-23-105850,microsoft_2023_ai_procurement}.

    \item \emph{Supply-chain security and transitive risk increase.}
    AI systems depend on upstream artifacts such as pre-trained models, datasets, and specialized toolchains. Unlike traditional software, these dependencies include model weights that can propagate backdoor vulnerabilities or data-poisoning risks into fielded capabilities. This necessitates provenance, integrity verification, and a more granular AI Bill of Materials (AIBOM) that extends traditional software provenance to include dataset lineage and model training configurations~\cite{intoto2019,slsa_v1_2023,sigstore2021,jiang2023pretrainedreuse,nist_ai_rmf_1_0}.

    \item \emph{The ecosystem changes quickly.}
    AI models, tools, and deployment stacks evolve rapidly, resetting baselines and increasing the need for continuous validation, monitoring, and adaptation across the lifecycle~\cite{aiindex2025,tfx_mlopsguide_2024,nist_ai_rmf_1_0}.

    \item \emph{Compute and hardware dependencies are often mission-relevant.}
    Many AI workloads depend on accelerator-rich compute environments such as GPUs, TPUs, or NPUs. As a result, performance, latency, throughput, and cost may vary significantly across hardware targets, including edge and cloud deployments~\cite{jouppi2017tpu,DLAccelSurvey2025,Purvish_24}.
\end{itemize}

\paragraph{Summary.}
These properties motivate the later scenario design and coding framework. In particular, they highlight why AI acquisition places greater pressure on data governance, probabilistic TEVV, lifecycle monitoring, provenance, and human-governance mechanisms than conventional software-intensive acquisition.

\subsection{Prior Work on SWP and AI Acquisition} \label{prior_work}
The DoD’s transition to AAF and SWP reflects a broader effort to support iterative delivery for mission-tailored software. Existing literature, however, remains limited and divided. One stream focuses on the institutionalization of Agile and DevSecOps, while another treats AI as a distinct governance and contracting problem.

The first stream of literature treats the SWP primarily as a reform mechanism for improving software-delivery speed. GAO found that, despite policy updates intended to support modern software practices, implementation remained inconsistent across programs \cite{gao_software_reform_2021}. Dunlap likewise emphasizes entrenched business practices, legacy certification processes, and workforce constraints as persistent barriers to software-acquisition reform \cite{dunlap_good_bad_ugly_2024}. Tate and Bailey further argue that SWP suitability depends on program conditions and organizational context rather than serving as a universal solution \cite{tate_bailey_feasible_2022}. While these studies identify constraints on SWP adoption for general software, they offer limited insight into whether these artifacts can accommodate AI acquisition.

The second stream of literature shifts attention to AI as a distinct acquisition and contracting problem driven by data dependence and continuous model evolution. The Center for Strategic and International Studies (CSIS)~\cite{csis2023pathforward} argues that procurement mechanisms designed for static software deliverables are ill-suited for AI systems requiring ongoing, data-driven retraining. Complementing this, DAU’s analysis~\cite{dau2022aiacquisition} advocates for a responsible AI framework that embeds ethics, transparency, and bias mitigation into oversight. These analyses provide a vision for what AI acquisition should entail, yet they remain largely decoupled from the concrete execution mechanisms of the SWP.

Recent DoD fielding efforts further complicate this landscape by introducing two diverging acquisition patterns. On one hand, generative AI is increasingly delivered through enterprise-scale commercial integration—such as the GenAI.mil rollout and large-scale Palantir awards—using license- and platform-centered contracting vehicles~\cite{google2025genaimil,army2025palantirEA}. 
Conversely, certain mission contexts, including forward operating positions, space systems, embedded systems, and contested environments, will require specialized, edge-based AI resident on mission systems and designed for real-time operation under tactical constraints~\cite{xtechOverwatch2025,DoDFY2026SOCOM,armyPEODemand2025}.  In those settings, AI acquisition aligns more closely with traditional program responsibilities, with the program office responsible for integration, sustainment, verification, and upgrade planning for AI components deployed on mission systems.

Prior research therefore, supports both SWP-driven modernization and AI-specific governance reform, but it says less about how those two agendas meet in practice. Despite rising AI funding and DoD’s reliance on SWP, relatively little work has examined whether SWP’s planning artifacts provide the actionable specificity required for AI-enabled systems, especially in tactical edge environments [6]. This paper addresses that gap through a scenario-based policy assessment that maps SWP planning artifacts to AI technical requirements, identifying key points of tension and proposing refinements for an AI-ready acquisition pathway.

\section{Research Question, Methodology, and Corpus Selection}

AI-enabled capabilities inherit many features of software-intensive systems, but they also introduce distinctive dependencies. Given these differences, it is crucial to understand whether the Software Acquisition Pathway (SWP) provides sufficiently specific guidance for acquiring such capabilities. We therefore investigate the following research question:

\begin{enumerate}
\renewcommand{\labelenumi}{\textbf{RQ:}}
  \item \emph{To what extent does the Software Acquisition Pathway (SWP), as implemented through the broader SWP-centered governance stack, accommodate AI’s distinctive technical and operational characteristics, and where does it remain underspecified?}

\end{enumerate}

\subsection{Methodology Overview}
\label{sec:method_overview}

Evaluating acquisition policy through observed program outcomes is difficult because such studies require long time horizons and are strongly shaped by local tailoring, organizational capacity, and operational context. They also depend on access to program artifacts, internal documentation, and personnel across multiple programs, which may be limited or unavailable in practice. In this setting, a lighter-weight pre-implementation assessment can provide an initial evidence base for whether a policy appears actionable and whether a more resource-intensive evaluation is warranted. We therefore employ \textbf{Policy Scenario Analysis (PSA)} \cite{cunningham2016_psa}. PSA evaluates policy through scenarios that combine relevant \emph{circumstances} with the \emph{policy actions} available to decision-makers, enabling analysts to examine likely implications before real-world execution \cite{cunningham2016_psa}. In this study, the documents described in Section~\ref{subsec:dod_document_environment} constitute the \emph{policy action set}. These documents define artifacts, roles, responsibilities, decision points, and review structures through which an SWP program is expected to plan and govern its activities. We operationalize PSA through a structured scenario-based assessment of a synthetic but grounded DoD-like AI acquisition case, alongside a conventional software comparator. In this way, the analysis helps surface weaknesses in the governing policy stack before they emerge in execution or sustainment.

\subsection{Scenario Design and Scope}
This subsection defines the baseline and comparator scenarios, the study boundary, and the stakeholder perspective used to evaluate policy actionability.

\paragraph{Scenario set design (baseline + comparator).}
PSA typically employs a small number of scenarios, usually beginning with a baseline case and comparing it with a single alternative; additional scenarios are introduced only when analytically necessary to clarify policy implications \cite{cunningham2016_psa}. In our assessment, we use two scenarios: (i) a baseline \emph{AI acquisition under SWP}, and (ii) a comparator \emph{conventional software-intensive acquisition under SWP}. The comparator serves as a methodological control that holds the acquisition framework constant while varying the technical character of the capability. This design allows us to distinguish AI-specific underspecification from broader SWP reliance on local tailoring.

\begin{figure}[t]
    \centering
    \begin{subfigure}{0.65\textwidth}
        \centering
        \includegraphics[width=\linewidth]{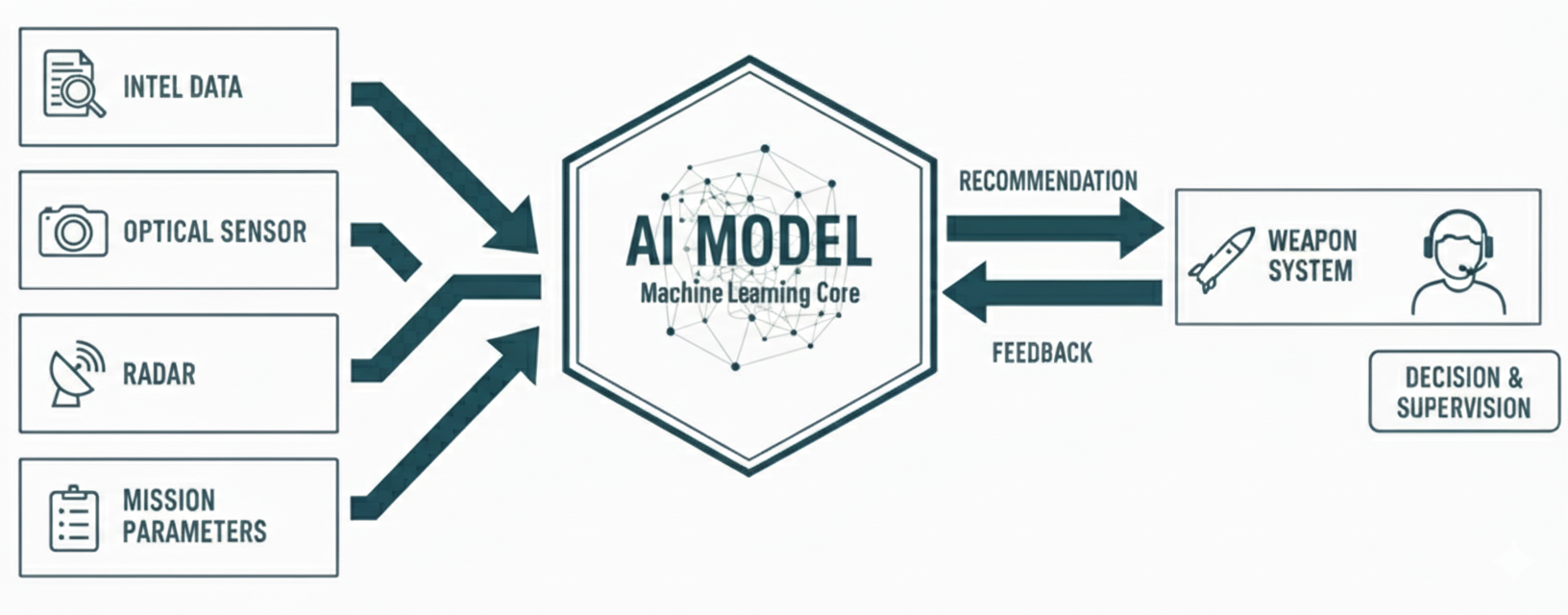}
        \caption{Notional AI-enabled targeting capability.}
        \Description{A flowchart showing sensor data, INTEL and mission parameters flowing into an AI inference engine, which then outputs targeting options to a human operator.}
        \label{fig:ai_model}
    \end{subfigure}

    \vspace{0.5em}

    \begin{subfigure}{0.65\textwidth}
        \centering
        \includegraphics[width=\linewidth]{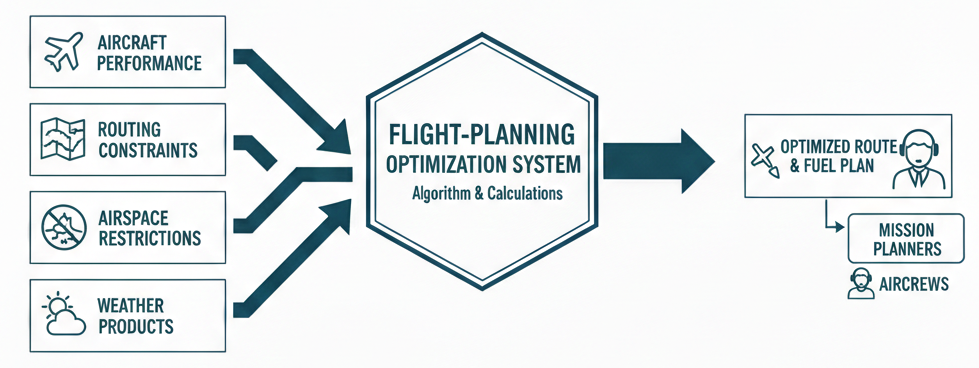}
        \caption{Notional flight planning optimizer.}
        \Description{A flowchart showing data sources ( aircraft performance, routing constraints, weather and restrictions) flowing into an optimization algorithm, which then outputs flight plans to a human operator.}
        \label{fig:flight_plan}
    \end{subfigure}

    \caption{Comparison of the two notional scenarios used in this study. (a) The AI-enabled targeting capability integrates diverse data streams---including signals intelligence (INTEL), optical sensors, radar, and mission parameters---into an AI inference engine that generates targeting recommendations for a weapon system, while a human operator remains in the loop for final decision authority and supervision. (b) The flight planning optimizer uses authoritative data sets such as aircraft performance specifications, airspace restrictions, and weather products within an optimization algorithm to produce a fuel-efficient, constraint-satisfying route for review by mission planners before transmission to aircrews.}
    \label{fig:scenario_models}
\end{figure}

\paragraph{Baseline scenario design: AI-enabled program (Targeting AI)}
Our primary evaluation case is a hypothetical capability termed the \emph{Targeting AI} program. The scenario is intentionally synthetic but grounded: rather than reproducing any single real-world program, it represents a recognizable class of DoD AI acquisition efforts in which machine-learning models are integrated into mission workflows under operational and governance constraints. The notional system fuses multi-source sensor data, including optical imagery, radar, and signals intelligence (SIGINT) reports, to generate targeting recommendations for human-in-the-loop decision support (see~\cref{fig:ai_model}). We assume that the capability must operate at the tactical edge, requiring on-premises execution in bandwidth-constrained and contested environments. This assumption is consistent with current DoD demand signals indicating that, alongside enterprise AI platforms, some mission contexts will require specialized, edge-based AI resident on mission systems\footnote{See \emph{DoD FY2026 Budget Estimates, SOCOM} (2025): ``Investments in 'AI at the Edge' are critical for autonomous ISR platforms operating in bandwidth-constrained and GPS-denied environments, where centralized AI processing is not a viable COA [Course of Action].''}  and operating under real-time tactical constraints~\cite{xtechOverwatch2025,DoDFY2026SOCOM,armyPEODemand2025}.

We further ground the scenario by mapping its principal features to publicly documented DoD program patterns, each of which contributes an acquisition-relevant burden preserved in the baseline.
\begin{itemize}[leftmargin=2em]
\item \emph{Analytic function:} Informed by \emph{Project Maven}, the scenario retains the requirement to process heterogeneous intelligence inputs using computer vision and machine learning in support of faster analysis and human decision-making~\cite{ProjectMavenMemo2017}.
\item \emph{Operational integration:} Informed by the \emph{PM IS\&A} portfolio, the system is situated within existing expeditionary intelligence-processing workflows, ensuring that the software must interoperate with mission systems and support tactical information requirements~\cite{PMISAWeb}.
\item \emph{Edge constraints:} Informed by current demand signals associated with \emph{CJADC2} and contested operations, the scenario assumes disconnected or intermittent connectivity, making update timing, local model validation, and sustainment planning acquisition-relevant burdens rather than merely architectural details~\cite{DoDC3Strategy2020}.
\end{itemize}

We design the \emph{Targeting AI} program scenario to preserve several of the AI-relevant acquisition properties summarized in~\cref{tab:ai_properties_dimensions} while remaining sufficiently bounded for structured policy analysis. It therefore serves as a demanding but analytically tractable test case: multi-source fusion makes data a first-class deliverable; environmental and mission drift create a need for monitoring and retraining to manage model evolution; the non-deterministic character of AI complicates conventional test logic; and reliance on specialized hardware at the edge makes deployment context an acquisition-facing concern.

\paragraph{Comparator scenario design: conventional software program (Flight Planning Optimization).}
To distinguish AI-specific underspecification from broader SWP reliance on local tailoring, we define a conventional software-intensive comparator with a similar operational context but without learning-based behavior: a notional \emph{flight planning optimization} application that generates fuel-efficient flight plans by integrating deterministic data sources and operational constraints such as aircraft performance models, routing constraints, and weather services, with mission-planning users reviewing outputs before dissemination to aircrews (see~\cref{fig:flight_plan}). As with the \emph{Targeting AI} baseline, this scenario is synthetic but grounded in publicly described DoD mission-planning software with comparable operational context, delivery pressures, and user interaction requirements~\cite{dote2018_mps_jmps_af}. We trace it through the same SWP decision points and planning artifacts as the AI scenario so that the acquisition framework remains constant while the technical character of the capability varies. The comparator therefore serves as a methodological control, helping us distinguish AI-specific acquisition burdens from broader forms of SWP underspecification that may also affect conventional software.

\paragraph{Scope and time horizon.}
We bound the study to the SWP Planning Phase, beginning after AAF pathway selection and continuing through completion of the planning activities required under SWP (\cref{synthesis_swp}). We assume that SWP has been formally designated as the program’s primary governance mechanism, consistent with Department memoranda directing its use for software-dominant capabilities \cite{LaPlante2022SWPDBS, DoD2025SoftwareAcquisition}. Within this boundary, we evaluate the guidance used to develop the required planning artifacts. We focus on the Planning Phase because it is the point at which written policy is converted into executable program governance. This makes it possible to assess whether the SWP-centered governance stack provides acquisition teams with sufficient direction before the program enters resource-intensive development and deployment.

\paragraph{Stakeholders.}
The PSA is conducted from the perspective of the Integrated Product Team (IPT), which is responsible for translating SWP policy into executable plans, artifacts, and reviewable evidence. Core IPT roles include the Program Manager and Lead Engineer, supported by functional experts in TEVV, cybersecurity, and contracting. The Decision Authority provides pathway and gate approvals and accepts the program’s risk posture, while functional authorities influence the minimum evidence required for release decisions. We adopt this perspective because SWP’s actionability for AI depends on whether program teams are given sufficiently clear guidance to execute consistently across iterations.

\subsection{Evaluation Procedure}
This subsection describes the procedure used to evaluate the study scenarios, including the episode structure, assessment criteria, coding scheme, and resulting analytic outputs.

\begin{figure}
  \includegraphics[width=\linewidth]{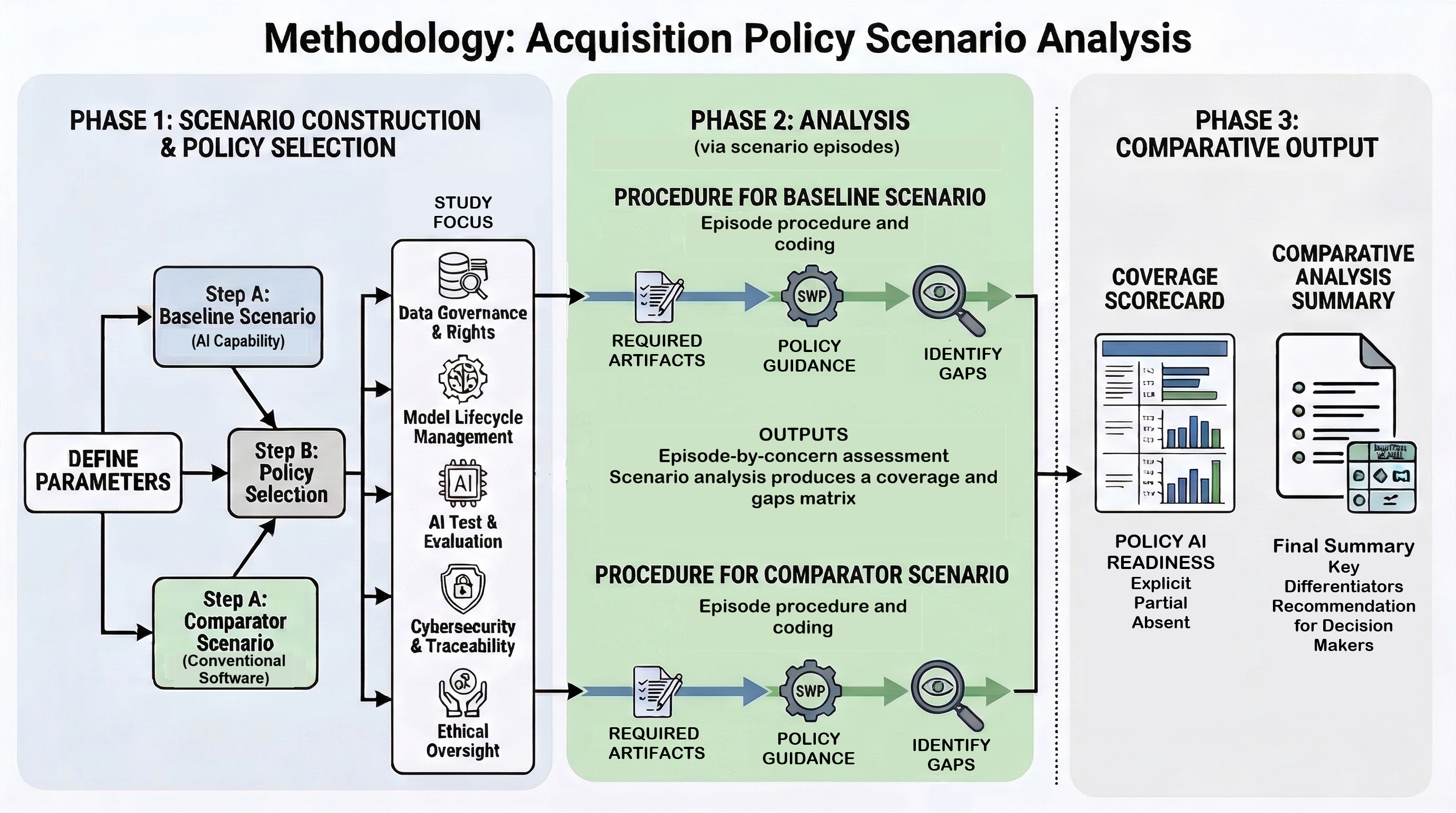}
  \caption{ Acquisition Policy Scenario Analysis methodology. We apply Policy Scenario Analysis to evaluate SWP actionability for AI by (Phase 1) defining study parameters, selecting the governing policy corpus, and constructing a baseline AI scenario plus a conventional-software comparator; (Phase 2) walking each scenario through SWP Planning-phase activities to map required artifacts to policy guidance and identify gaps; and (Phase 3) producing comparative outputs—a coverage scorecard and a summary of key differentiators and recommendations for decision makers.}
  \Description{A three-phase horizontal flow diagram depicting the Policy Scenario Analysis (PSA) workflow. Phase 1, on the left, shows input boxes for the 'Policy Corpus' and 'Scenario Definition' (AI vs. Conventional). Phase 2, in the center, illustrates the analytical engine where scenarios are mapped against SWP Planning activities to identify policy gaps. Phase 3, on the right, displays the final outputs as a 'Coverage Scorecard' and a 'Recommendations Report' for decision makers.}
  \label{fig:Method}
\end{figure}

\paragraph{Scenario Episodes (Circumstances).}
To instantiate the \emph{circumstances} side of the PSA framework, we organize our scenarios into five episodes: (1) data sourcing, labeling, and rights negotiation; (2) TEVV design; (3) cybersecurity and release provenance for model and software artifacts; (4) lifecycle management; and (5) human oversight constraints for operational use. We initially identified ten candidate episodes from the intersection of three elements in the study design: the AI-relevant acquisition properties identified in~\cref{tab:ai_properties_dimensions}, the mandatory SWP Planning Phase artifacts and review structures described in the governing corpus, and the recurring acquisition burdens identified in the AI-governance and defense-acquisition literature reviewed in~\cref{prior_work}. From that broader set, we selected five because they provided the strongest non-redundant coverage of AI-relevant stress points within the SWP Planning Phase. We do not claim that these five episodes exhaust all possible AI-acquisition circumstances; instead, they were chosen as representative and analytically useful cases for evaluating the SWP-centered governance stack.

\paragraph{Evaluation concerns.}
Within each episode, we assess whether SWP guidance provides actionable policy support for five acquisition concerns that are especially salient for AI and that can also be examined in the conventional software comparator:
\begin{itemize}[leftmargin=2em]
    \item \emph{Data governance and data rights:} whether policy supports ownership and accountability, provenance and labeling expectations, and rights sufficient for training, test, operation, and sustainment.
    \item \emph{Assurance and T\&E (TEVV):} whether policy supports the evidence and evaluation methods needed to justify performance across operational conditions, including deterministic software verification as well as more condition-sensitive or non-deterministic AI behavior.
    \item \emph{Cybersecurity and traceability:} whether policy supports model and pipeline security (including supply chain) and traceability across data, model, software, and system releases.
    \item \emph{Lifecycle management:} whether policy supports monitoring, update and change-control responsibilities, and triggers for retraining, drift detection, software revision, or re-authorization following material changes in data, models, code, or operational conditions.
    \item \emph{Human oversight:} whether policy supports human oversight and responsible or ethical compliance mechanisms appropriate to the use context.
\end{itemize}

These same concern areas are applied to the conventional software comparator, but they arise in different forms and with different levels of acquisition burden. For example, data governance in the comparator primarily concerns authoritative input data and interface control rather than training-data rights; lifecycle management centers on software versioning and rule updates rather than retraining or model drift; and assurance focuses on deterministic verification and test coverage rather than stochastic model behavior. Applying a common concern structure across both scenarios allows observed differences in policy support to be attributed more clearly to AI-specific properties rather than to differences in the evaluation framework. By selecting these five areas, we operationalize acquisition actionability for AI in a manner consistent with the DoD Responsible AI Strategy’s tenets of being \emph{Traceable}, \emph{Reliable}, and \emph{Governable} \cite{DoD_RAI_SIP_2022}.

\paragraph{Episode Procedure and Coding.}
To implement PSA in a repeatable manner, we developed a structured episode worksheet~\cite{miles_2014}. This instrument records the specific context of an episode, including: circumstances, assumptions, policy actions, governing corpus evidence, coverage judgments, gap types, and downstream consequences~(\cref{fig:psa_worksheet}).  It operationalizes the scenario-based logic of PSA and supports consistent cross-episode comparison. The unit of analysis is the episode-by-concern assessment of whether the applicable policy action set provides actionable support through required artifacts, minimum content expectations, or decision gates. Each episode is analyzed in three steps: (1) specify the circumstances and assumptions; (2) identify the SWP policy actions invoked and extract the associated artifact content from the governing corpus; and (3) evaluate the actionability of that content for the concern under review.

\begin{figure}[t]
\centering
\footnotesize
\fbox{
\begin{minipage}{0.8\columnwidth}
\textbf{Example Episode Worksheet}

\vspace{0.5em}
\textbf{Episode / concern:} Data governance and data rights for the Targeting AI program / whether policy supports ownership and accountability, provenance and labeling
expectations, and rights sufficient for training, test, operation, and sustainment

\vspace{0.5em}
\textbf{Coverage judgment:} Partial

\vspace{0.5em}
\textbf{Gap type:} Missing artifact content

\vspace{0.5em}
\textbf{Circumstances and assumptions:}

\vspace{0.5em}
The program must acquire, label, protect, and sustain multi-source data for training, validation, and operation, with future retraining and sustainment expected.

\vspace{0.5em}
\textbf{Policy actions invoked:}

\vspace{0.5em}
Planning Phase artifacts include the Program Protection Plan, IP Strategy, and Data Strategy within the Acquisition Strategy.

\vspace{0.5em}
\begin{itemize}[leftmargin=1.2em,noitemsep,topsep=0pt]
    \item DoDI~5000.83 supports protection of controlled technical information and data assets, helping the program define how sensitive training and operational data should be safeguarded and governed.
    \item DoDI~5010.44 supports negotiation of data and IP rights, helping the program address ownership, access, and reuse rights for the data needed for training, testing, operation, and later retraining.
    \item DoDI~5000.82 and the DoD Data, Analytics, and AI Adoption Strategy support enterprise data governance, helping frame expectations for stewardship, accountability, and management of data across the AI lifecycle.
\end{itemize}

\vspace{0.3em}
Coverage is partial because these sources support protection, rights-planning, and governance at a general level, but do not specify program-level requirements for data provenance, labeling quality, quality-assurance thresholds, or retraining-ready data deliverables.

\vspace{0.5em}
\textbf{Downstream consequence:}

\vspace{0.5em}
Data may be protected and rights partially secured, yet provenance, labeling quality assurance, versioning, and retraining-ready deliverables remain insufficiently specified.

\end{minipage}
}
\caption{Episode worksheet developed for this study to operationalize Policy Scenario Analysis in the software acquisition context. The example shows how the data-governance and data-rights episode was coded.}
\Description{A framed worksheet titled 'Example Episode Worksheet' used for qualitative policy analysis. It is organized into six structured sections: Episode/Concern (Data governance for Targeting AI), Circumstances (multi-source data requirements), Policy Actions (specific strategies invoked), Governing Corpus Evidence (citations of DoDI 5000.83, 5010.44, and 5000.82), a Coverage Judgment (Partial), and Downstream Consequences (lack of provenance and retraining-ready deliverables).}
\label{fig:psa_worksheet}
\end{figure}

Following Cunningham’s observation that policy failure often stems from underspecification rather than a total absence of guidance~\cite{cunningham2016_psa}, we evaluate actionability using a ternary coding scheme: \emph{Explicit}, \emph{Partial}, and \emph{Absent}. This scheme distinguishes between guidance that is directly actionable, guidance that is relevant but insufficiently specified for repeatable execution, and guidance that is missing at the point of decision. It also aligns with qualitative approaches that distinguish degrees of support or maturity rather than treating adequacy as purely binary~\cite{ragin2008redesigning,cmmi2024_levels}. We define these codes as follows:

\begin{itemize}[leftmargin=2em]
    \item \emph{Explicit:} the policy and/or required artifact includes a direct mandate, role assignment, decision trigger, or specific content requirement addressing the property.
    \item \emph{Partial:} the policy provides a relevant artifact or general software guidance, but lacks the technical specificity, standards, or triggers needed to manage the risk in a repeatable way.
    \item \emph{Absent:} the policy corpus provides no meaningful guidance for the property at that decision point, leaving resolution to local tailoring or undocumented practice.
\end{itemize}

Codes are assigned conservatively based on the most specific applicable policy language in the governing corpus for the episode and concern under review. In this paper, an \emph{Explicit} rating does not require that every element of support appear in core SWP guidance alone; rather, it requires that the applicable SWP-centered governance stack provide direct and actionable support for the property at issue. When coverage is \emph{Partial} or \emph{Absent}, we additionally classify the gap type (missing artifact content, unclear responsibility assignment, missing decision gate or trigger, or missing contracting or rights mechanism) and record a brief downstream consequence describing the likely execution or sustainment risk if the gap persists.

\paragraph{Answering the research question and outputs.}
We answer the research question by using the episode worksheet to determine whether the governing corpus provides program-actionable guidance for AI-enabled acquisition and to identify where the SWP-centered governance stack remains underspecified. We aggregate episode-level judgments into two outputs: (i) a scenario-to-policy mapping linking SWP steps to governing provisions, required artifacts, and responsible roles or authorities; and (ii) a coverage-and-gaps matrix summarizing Explicit, Partial, and Absent judgments and associated gap types across the artifact-producing guidance in the corpus. Together, these outputs show how well SWP accommodates AI’s distinctive characteristics and where additional specificity is required for consistent execution across program offices.

\paragraph{Completeness and stopping criteria.}
Consistent with PSA, we treat scenarios as decision-support instruments rather than literal predictions and therefore use a small number of representative scenarios designed to reveal hidden weaknesses and their consequences \cite{cunningham2016_psa}. We consider the assessment complete when (i) all mandatory SWP Planning Phase artifacts, decision points, and phase entry or exit criteria within scope have been enumerated; (ii) each in-scope AI-relevant property is exercised by at least one episode such that it must map to a policy-backed artifact, role, or decision gate; and (iii) further passes through the governing corpus do not introduce additional artifacts, responsibilities, or decision mechanisms that would change the existing mappings. Following scenario-based evaluation guidance, we do not attempt to enumerate all possible operational states. Instead, we select episodes to cover representative conditions the program is likely to encounter and to expose where guidance is technology-agnostic or absent~\cite{looker2008scenario}.

\paragraph{Method tailoring.}
Our implementation adapts PSA to the constraints and goals of this study. Rather than constructing alternative futures or using expert elicitation to assign future probabilities, we use a baseline/comparator scenario design and a structured episode worksheet to trace how the applicable policy guidance is translated into roles, artifacts, and gates. This tailoring preserves PSA’s core purpose of surfacing hidden weaknesses and anticipating their consequences, while producing audit-ready, program-actionable findings for acquisition practice \cite{cunningham2016_psa}.

\subsection{Limitations and Threats to Validity}
We discuss construct, internal, and external threats to validity as they apply to this scenario-based policy assessment. Because PSA is a decision-support method rather than a predictive evaluation of observed program outcomes, these threats primarily concern how we operationalize acquisition actionability, how we draw inferences from scenario analysis, and how far the findings can reasonably be generalized beyond the study setting.

\paragraph{Construct validity.}
Our central construct is acquisition \emph{actionability}: whether the SWP-centered governance stack gives a program office sufficiently clear roles, artifacts, content expectations, and decision triggers to manage AI-relevant acquisition burdens. We operationalize this construct through five evaluation concerns and an episode-based PSA worksheet focused on Planning Phase artifacts and gates. This framing is appropriate to our research question, but it does not capture every possible dimension of AI governance or acquisition success. In particular, our study evaluates policy support for planning and governance rather than downstream operational effectiveness, contractor performance, or organizational adoption in practice.

\paragraph{Internal validity.}
Our analysis relies on qualitative judgments when mapping policy language to scenario episodes and when assigning \emph{Explicit}, \emph{Partial}, or \emph{Absent} codes. The coding was performed by a single analyst, and we therefore do not report inter-rater agreement. This reflects the specialized nature of the assessment: it required familiarity with DoD acquisition policy, pathway artifacts, and military acquisition practice, and one author contributed substantially more domain expertise in those areas than the other. We mitigate this limitation through a structured worksheet, explicit code definitions, a bounded Planning Phase scope, and a baseline/comparator design that holds the acquisition framework constant across scenarios. Even so, the findings should be read as structured analytic judgments rather than coder-independent measurements. We discuss this division of expertise further in the positionality statement below~(\cref{subsec:positionality}).

\paragraph{External validity.}
Our results are derived from a synthetic but grounded DoD-like AI baseline, a conventional software comparator, and a governing corpus centered on SWP and associated guidance. The findings are therefore most applicable to software-intensive defense programs in which AI-enabled capabilities have similar deployment requirements. They may be less applicable to simpler enterprise AI deployments such as GenAI.mil (\cref{prior_work}), autonomous systems outside the scope of our baseline, or acquisition environments outside the DoD. More broadly, PSA does not aim to predict the outcome of any single real program~\cite{cunningham2016_psa}; rather, it provides analytically useful insight into where the current SWP-centered governance stack is likely to be actionable, underspecified, or dependent on undocumented local practice.


\subsection{Statement of Positionality}
\label{subsec:positionality}
The research team comprises an acquisition professional and a software engineering researcher. This dual perspective informed the study’s design by balancing methodological rigor with operational realism. Our combined expertise led us to prioritize software-centric guidance as the primary analytical lens and ensured that the selected scenarios reflect the practical constraints of the defense acquisition environment.

Because the acquisition-policy coding required familiarity with DoD acquisition pathways, artifacts, and practical application, the primary coding judgments were made solely by the author with direct defense-acquisition experience. The findings and interpretations presented herein are strictly those of the authors and do not represent the official positions of any government agency, military department, or commercial vendor.

\section{Results}
\label{sec:analysis}

This section presents the episode-level results of applying PSA to the baseline \emph{Targeting AI} scenario through the structured episode worksheet developed for this study (\cref{fig:psa_worksheet}). For each episode, we assess how the SWP-centered governance stack addresses the relevant acquisition concern, identify remaining gaps and their likely consequences, and use the conventional software comparator to distinguish AI-specific underspecification from broader limitations in SWP guidance.

\subsection{Summary of Baseline Scenario Findings}
Table~\ref{tab:baseline_summary} summarizes the PSA results for the baseline \emph{Targeting AI} scenario. Overall, the SWP-centered governance stack provides a usable acquisition structure but uneven AI-specific actionability. The strongest support appears in AI assurance and TEVV, where SWP artifacts and supplemental CDAO guidance provide relatively direct direction for planning evaluation activities. In the remaining episodes, the guidance provides a workable structure but leaves important elements of artifact content, review triggers, and decision criteria to local interpretation.

\begin{table*}[t]
\centering
\small
\caption{Summary of PSA baseline findings for the \emph{Targeting AI} scenario, assessed at the level of the SWP-centered governance stack.}
\label{tab:baseline_summary}
\begin{tabular}{p{2.9cm} p{1.5cm} p{4.0cm} p{5.1cm}}
\toprule
\textbf{Episode / concern} & \textbf{Coverage} & \textbf{Primary artifact(s)} & \textbf{Key finding} \\
\midrule
Data governance and data rights (Illustrated Fig.~\ref{fig:psa_worksheet})
& Partial
& Program Protection Plan; IP Strategy; Data Strategy within Acquisition Strategy
& Governance stack provides explicit mechanisms for data protection and rights-planning, but does not specify minimum content for provenance, labeling quality, versioning, or the data deliverables needed for retraining and sustainment. (Gap type: missing artifact content)
\\
\addlinespace

AI assurance and TEVV
& Explicit
& Test Strategy; T\&E planning artifacts; CDAO AI T\&E guidance
& Governance stack provides explicit support for AI-specific TEVV when the required Test Strategy is combined with CDAO guidance on stochastic performance criteria, operational-envelope characterization, and other AI-specific evaluation factors.
\\
\addlinespace

Cybersecurity and traceability
& Partial
& Cybersecurity Strategy within Acquisition Strategy; Program Protection Plan; Cybersecurity Strategy Annex
& Governance stack is strong on lifecycle cybersecurity, supply-chain risk management, and recurring assessment, but remains underspecified for end-to-end provenance linking datasets, model artifacts, software builds, dependencies, and deployed system configurations. (Gap type: missing artifact content)
\\
\addlinespace

Lifecycle management
& Partial
& Product Support Strategy; lifecycle-support content within the Acquisition Strategy; OMB M-25-22
& Governance stack provides meaningful support for post-award monitoring and update governance, but drift detection, retraining triggers, and AI-specific performance thresholds remain only partially specified. (Gap type: missing decision gate/trigger)
\\
\addlinespace

Human oversight
& Partial
& Acquisition Strategy; DoD Responsible AI Strategy and Implementation Pathway; human-systems planning artifacts
& Governance stack provides meaningful guidance on accountability, intended-use constraints, and governability, but does not yet embed these expectations in pathway-specific artifact sections, review gates, or standard decision criteria. (Gap type: missing decision gate/trigger)
\\
\bottomrule
\end{tabular}
\end{table*}

\subsection{Data Governance and Data Rights}
\label{subsec:data_rights}
\textbf{Finding: }Coverage is \emph{Partial} because the governance stack fails to translate enterprise quality standards into program-level expectations for training data provenance, labeling QA, or retraining-ready data deliverables.
\subsubsection{Baseline Scenario: Targeting AI}
\paragraph{Circumstance.}
In this episode, the \emph{Targeting AI} program must acquire, prepare, and sustain the data needed to train, validate, and operate a targeting model that fuses multiple sensor sources and provides targeting recommendations. The scenario assumes that the data requires labeling and quality assurance before use, and must remain available for later retraining, test, and sustainment as mission conditions evolve. These circumstances force the program to address acquisition-time questions of dataset provenance, labeling integrity, version control, protection of sensitive data products, and government rights to use, modify, and sustain data and associated artifacts over time.

\paragraph{Policy Actions Invoked.}
Under the SWP-centered governance stack, the IPT is required to develop a set of planning artifacts that translate pathway guidance into executable program decisions. In this episode, the circumstance brings three required artifacts directly into play: the Program Protection Plan (PPP), the Intellectual Property (IP) Strategy, and the program’s Data Strategy within the Acquisition Strategy. These artifacts are therefore the principal policy actions through which the IPT must address the episode’s data-governance and data-rights issues. The governing documents implicated by this episode include DoDI~5000.83 for program protection, DoDI~5010.44 for intellectual property and data rights, and DoDI~5000.82 together with the DoD Data, Analytics, and AI Adoption Strategy for enterprise data governance \cite{DoDI5000_83,DoDI5010_44,DoDI5000_82,AIAdoption2023}.

\paragraph{Assessment.}
At the episode-by-concern level, coverage for data governance and data rights is \emph{Partial}. When evaluated against our concern whether policy supports ownership and accountability, provenance and labeling expectations, and rights sufficient for training, test, operation, and sustainment, the governance stack exhibits uneven actionability. 

The PPP, governed primarily by DoDI~5000.83, provides an explicit mechanism for identifying and protecting controlled technical information and mission-critical program information across the lifecycle \cite{DoDI5000_83}. In the Targeting AI case, this mechanism can be applied to training datasets, model weights, and feature pipelines as sensitive program assets. However, the guidance is oriented toward preventing exfiltration or compromise rather than ensuring the usability, integrity, and technical reliability of AI data. It therefore provides an explicit basis for protection, but only indirect support for broader AI data governance.

The IP Strategy provides more direct support. DoDI~5010.44 explicitly directs programs to plan early for the intellectual property and data rights needed to keep systems functional, affordable, and supportable over time \cite{DoDI5010_44}. For the \emph{Targeting AI} program, this gives the IPT a clear policy basis to identify and negotiate rights for training datasets, documentation, model-supporting technical data, and related deliverables needed for sustainment. In that sense, coverage is explicit: the policy clearly requires early planning for the rights the government will need to use, modify, and support the capability over time.

The program’s Data Strategy, developed as part of the Acquisition Strategy, draws on DoDI~5000.82 and the DoD Data, Analytics, and AI Adoption Strategy to provide a meaningful enterprise-level basis for data governance. In particular, this guidance treats data as a product, emphasizes lifecycle assessment using VAULTIS (\emph{Visible, Accessible, Understandable, Linked, Trustworthy, Interoperable, and Secure}) \cite{DoDI5000_82,AIAdoption2023} and related data-quality dimensions, and assigns responsibility to data domain owners and data product teams. Despite these high-level principles, coverage is coded \emph{Partial} for program-actionable acquisition guidance because provenance and labeling expectations are left to local implementation and are not translated into standard minimum content for SWP planning artifacts. Consequently, the pathway currently lacks the technical specificity needed to manage these AI-relevant properties in a repeatable, actionable way.

\paragraph{Gap and Downstream Consequence.}
The gap in this episode is \emph{missing artifact content}. The governing documents do not clearly require planning artifacts to define training data provenance, labeling quality-assurance thresholds, or dataset versioning rules. As a result, a program may secure enough protection and rights to field an initial capability while still lacking the metadata and governance needed to sustain model performance. Without clear responsibility for producing and maintaining labels and dataset documentation, retraining may become infeasible in practice.

\subsubsection{Comparator Scenario: Flight Planning Optimizer}
\paragraph{Episode Summary and Assessment.}
The comparator episode involves a similar acquisition problem: ensuring that the flight-planning system has continuing access to the data it depends on, including aircraft performance data, weather products, airspace restrictions, and related mission-planning inputs. The same general planning artifacts are implicated, particularly the Program Protection Plan, the Intellectual Property Strategy, and the data-related portions of the Acquisition Strategy. In this case, however, the relevant inputs are established operational data products with designated owners, stewardship mechanisms, and update processes. As a result, the SWP-centered governance stack is generally sufficient to document access, protection, and sustainment arrangements without introducing the additional governance burden associated with training and evaluation datasets.

\paragraph{Comparative Interpretation.}
The comparator shows that SWP is generally adequate when a program depends on established operational data with known stewardship and update processes. The burden becomes more demanding in the \emph{Targeting AI} baseline because training and evaluation datasets must themselves be governed as enduring elements of the capability, including their provenance, labeling, reuse rights, and retraining suitability.

\subsection{AI Assurance and TEVV}
\label{subsec:tevv}
\textbf{Finding: }Coverage is \emph{Explicit} because the required SWP testing structure, when combined with CDAO AI T\&E guidance, directly supports operational-envelope characterization and AI-specific TEVV.
\subsubsection{Baseline Scenario: Targeting AI}
\paragraph{Circumstance.}
In this episode, the \emph{Targeting AI} program must define how the system will be tested, evaluated, verified, and validated before and during iterative delivery. The program must assess a model-based targeting capability whose behavior may vary across operational and environmental conditions, including cases in which performance degrades or fails unpredictably. These circumstances raise acquisition-time questions about what constitutes acceptable model performance, how to characterize behavior across an operational envelope, and what forms of evidence are sufficient to support release and continued operational use.

\paragraph{Policy Actions Invoked.}
Within the SWP-centered governance stack, the IPT is required to produce a Test Strategy during the Planning Phase. In this episode, the Test Strategy and related T\&E planning artifacts are therefore the principal documents through which the program must translate AI-assurance needs into executable test planning. The governing policy documents implicated by this episode are anchored in DoDI~5000.87, which establishes SWP testing expectations, and DoDI~5000.89, which provides the Department’s functional policy for test and evaluation and the framework used to develop the Test Strategy. They also include supplemental AI-specific T\&E guidance from CDAO that bears directly on how those expectations can be operationalized for AI-enabled systems \cite{DoDI5000_87,DoDI5000_89,CDAO_SI_TE_2024}.

\paragraph{Assessment.}
At the episode-by-concern level, coverage for AI assurance and TEVV is \emph{Explicit}. DoDI~5000.87 and DoDI~5000.89 provide the required planning structure and baseline software T\&E expectations, while supplemental CDAO AI T\&E guidance supplies the more specific evaluation content needed to assess non-deterministic behavior across operational conditions. Taken together, these documents provide direct support for defining acceptable model performance, characterizing behavior across an operational envelope, and specifying evidence appropriate for release and continued operational use.

DoDI~5000.87 describes the Test Strategy as the program’s blueprint for how capabilities, features, and user stories will be tested to satisfy developmental test and evaluation criteria and to demonstrate operational effectiveness, suitability, interoperability, survivability, and cyber survivability throughout the software lifecycle \cite{DoDI5000_87}. Likewise, DoDI~5000.89 emphasizes continuous integration, automated data collection, and reusable test scripts, tools, and libraries to support iterative evaluation \cite{DoDI5000_89}. However, these core SWP documents are less specific on AI-tailored acceptance logic, such as stochastic performance criteria, operational-envelope characterization, and evaluation under distributional or environmental variation.

Those more specific expectations are supplied by the supplemental CDAO Test and Evaluation Strategy Framework and companion AI T\&E guidance, which explicitly address AI-specific metrics, operational-condition characterization, and related evaluation factors in acquisition-relevant terms \cite{CDAO_SI_TE_2024}. As a result, coverage for AI TEVV is explicit when this supplemental guidance is incorporated in our governance stack.

\paragraph{Gap and Downstream Consequence.}
Because the SWP-centered governance stack provides explicit support for AI-specific TEVV when read as a whole, the weakness in this episode is incomplete integration of that guidance into core SWP testing expectations. The AI-specific content needed for strong TEVV planning is not fully embedded in the core SWP testing guidance itself. As a result, program offices that do not identify and apply the relevant supplemental guidance may still default to narrower or overly deterministic test logic, creating variability in implementation quality across programs.

\subsubsection{Comparator Scenario: Flight Planning Optimizer}
\paragraph{Episode Summary and Assessment.}
In the comparator episode, the program team must define how the flight-planning capability will be tested before and during iterative delivery. The program must verify and validate a deterministic optimization application operating over known software logic, authoritative data inputs, and representative mission-planning conditions. The same general T\&E artifacts are implicated, particularly the Test Strategy and related planning mechanisms under DoDI~5000.87 and DoDI~5000.89. Because the comparator system does not depend on learned model behavior, drift-sensitive inference, or stochastic outputs, the core SWP and functional T\&E guidance are generally sufficient to support regression testing, scenario-based evaluation, and system-level verification.

\paragraph{Comparative Interpretation.}
The comparator shows that SWP is generally adequate when the capability can be evaluated through established software-verification and mission-scenario methods. In the \emph{Targeting AI} baseline, acceptable performance must also be defined across stochastic behavior and varying operational conditions, which makes supplemental AI-specific TEVV guidance more important to consistent program execution.

\subsection{Cybersecurity and Traceability}
\label{subsec:cyber_traceability}
\textbf{Finding: }Coverage is \emph{Partial}; while lifecycle security is strong, the governance stack does not require release-level provenance linking datasets, models, dependencies, software builds, and deployed configurations.
\subsubsection{Baseline Scenario: Targeting AI}
\paragraph{Circumstance.}
In this episode, the \emph{Targeting AI} program must prepare to release and sustain an AI-enabled targeting capability in a contested environment while maintaining confidence in both its cybersecurity posture and the provenance of each deployed release. The delivered capability depends on an evolving combination of code, model artifacts, data products, dependencies, configuration settings, and deployment pipelines. These conditions raise acquisition-time questions about how to secure development and deployment environments, manage supply-chain risk, and preserve traceability from a fielded release back to the specific data, model, weights, software, and configuration elements used to produce it.

\paragraph{Policy Actions Invoked.}
Within the SWP-centered governance stack, the IPT is required to address cybersecurity during planning through the Cybersecurity Strategy section of the Acquisition Strategy and through the Program Protection Plan (PPP) and associated Cybersecurity Strategy Annex. DoDI~5000.87 establishes cybersecurity as a continuous lifecycle obligation and requires a risk-based approach to be embedded across design, infrastructure, development, test, integration, delivery, and operations \cite{DoDI5000_87}. It further requires recurring assessment of the supply chain, development environment, processes, and tools, together with continuous automated cybersecurity testing and operational evaluation. DoDI~5000.90 strengthens these requirements by assigning explicit PM accountability for cybersecurity across acquisition stages, requiring identification and documentation of breach consequences, and enforcing cybersecurity through RMF and supply-chain risk management (SCRM) mechanisms \cite{DoDI5000_90}. These artifacts and responsibilities are therefore the principal policy actions through which the IPT must address cybersecurity and release provenance in this episode.

\paragraph{Assessment.}
Coverage for cybersecurity and traceability is \emph{Partial}. When evaluated against our concern whether policy supports model and pipeline security (including supply chain) and traceability across data, model, software, and system releases, the governance stack exhibits uneven actionability. 

For cybersecurity, the guidance is comparatively strong. DoDI~5000.87 and DoDI~5000.90 clearly require programs to plan for cybersecurity across the lifecycle, assess development and supply-chain risks, and maintain continuous security monitoring and testing \cite{DoDI5000_87,DoDI5000_90}. In the \emph{Targeting AI} case, this gives the IPT a clear policy basis to secure the development pipeline, assess tools and dependencies, document breach consequences, and integrate cybersecurity into release planning and operational use.

The weaker area is traceability, particularly for AI supply-chain provenance. While DoDI~5000.87 mandates continuous cybersecurity practices, it does not translate these into AI-specific requirements for dataset integrity, model artifact protection, or vetting of third-party foundation models. Nor does it clearly require a release-level provenance artifact linking data versions, training configurations, model weights, dependencies, and deployed system configurations. In the \emph{Targeting AI} case, that omission creates a critical visibility gap: the IPT may be unable to determine whether a targeting failure stems from a code exploit, a data-poisoning attack, model compromise, or an untrusted upstream dependency. Coverage for AI-specific traceability and supply-chain provenance is coded as \emph{Partial}.

\paragraph{Gap and Downstream Consequence.}
The gap in this episode is \emph{missing artifact content}. The governing documents do not clearly require planning artifacts to define how dataset versions, training configurations, model artifacts, dependencies, and deployed system configurations will be linked and preserved across releases. If these linkages are not established during planning, the program may be able to secure and deploy the capability while still lacking the traceability needed to investigate failures, support forensic analysis, manage retraining-related supply-chain risk, or confidently roll back to a known-good configuration.

\subsubsection{Comparator Scenario: Flight Planning Optimizer}
\paragraph{Episode Summary and Assessment.}
In the comparator episode, the program office must secure the flight-planning capability and maintain confidence in released software versions. The program must protect conventional software builds, interfaces, dependencies, and operational data feeds and ensure that released versions can be traced through ordinary software configuration-management practices. The same policy actions are implicated, particularly the Cybersecurity Strategy, PPP, and associated RMF artifacts. Because the comparator system relies on authoritative data products rather than trained model artifacts, conventional configuration control and software-assurance practices are generally sufficient to preserve release provenance.

\paragraph{Comparative Interpretation.}
The comparator shows that SWP is generally adequate when release provenance can be maintained through conventional software configuration management and established cybersecurity controls. In the \emph{Targeting AI} baseline, the provenance problem extends across data, models, weights, dependencies, and deployment configurations, which makes the current lack of explicit AI-specific traceability expectations more consequential.

\subsection{Lifecycle Management}
\label{subsec:lifecycle_management}
\textbf{Finding: }Coverage for this episode is coded as \emph{Partial} because the governance stack does not embed clear drift-detection, retraining, and re-authorization triggers.
\subsubsection{Baseline Scenario: Targeting AI}
\paragraph{Circumstance.}
In this episode, the program team must plan for how the capability will be sustained, monitored, and updated after initial fielding. Operational performance may change over time as data distributions shift, environments evolve, and models are retrained or replaced. These conditions raise acquisition-time questions about who is responsible for post-deployment monitoring, what evidence should trigger retraining or update decisions, and how material model changes should be reviewed, approved, and re-authorized during iterative execution.

\paragraph{Policy Actions Invoked.}
Within the SWP-centered governance stack, the IPT is required to develop a Product Support Strategy during the Planning Phase to define how the capability will remain functional, supportable, and affordable across the lifecycle. In software acquisition practice, this artifact may appear as a stand-alone Product Support Strategy or as lifecycle-support content within the Acquisition Strategy \cite{dau_SW_acq_strategy_2022,dafi63-101_2024}. The governing policy documents implicated by this episode therefore include both the product-support guidance associated with SWP and the supplemental AI-relevant acquisition guidance that informs post-award AI governance. In particular, OMB Memorandum M-25-22 is included because it provides acquisition-relevant direction on post-award AI monitoring, evaluation, and update governance~\cite{omb_m25_22_2025}.

\paragraph{Assessment.}
Overall coverage for lifecycle management is \emph{Partial}. When evaluated against our concern whether policy supports monitoring, update and change-control responsibilities, and triggers for retraining, drift detection, or re-authorization following material changes, the governance stack exhibits uneven actionability.

The Product Support Strategy provides explicit mechanisms for the lifecycle sustainment of software systems, aligning with modern iterative practice by requiring plans for continuous delivery, software versioning, and long-term sustainment~\cite{dau_SW_acq_strategy_2022,dafi63-101_2024}. Furthermore, the governance stack is materially strengthened by OMB Memorandum M-25-22, which provides acquisition-relevant mechanisms for governing AI systems after award. It explicitly supports contractual terms for ongoing testing and monitoring, recommends use of agency-controlled validation or testing datasets representative of deployed conditions, and provides policy-backed hooks for vendor access, independent evaluation, update governance, rollback expectations, and performance standards that new versions must satisfy before deployment \cite{omb_m25_22_2025}. For the \emph{Targeting AI} scenario, these provisions make post-deployment monitoring and update governance sufficiently direct and actionable to justify an \emph{Explicit} sub-assessment within the broader episode.

The primary limitation for the episode is the lack of drift detection as a standardized technical control. While OMB M-25-22 mandates ongoing monitoring, it provides no baseline for identifying distributional shift or determining specific retraining thresholds. As such, the SWP-centered governance stack facilitates lifecycle administration but provides only \emph{Partial} support for the automated or statistical detection of performance degradation. 

\paragraph{Gap and Downstream Consequence.}
The primary shortfall in this episode is \emph{missing decision gates and triggers}. Because AI-specific monitoring and re-authorization expectations are not embedded in core SWP guidance, they must be assembled from a fragmented collection of supplemental documents. This creates a risk of procedural flattening, where the IPT may treat material model changes as ordinary software updates, bypassing the rigorous performance-integrity checks required for non-deterministic systems. Even with the full governance stack, the lack of standardized drift-detection criteria leaves program offices with persistent uncertainty regarding when declining operational performance necessitates formal intervention.

\subsubsection{Comparator Scenario: Flight Planning Optimizer}
\paragraph{Episode Summary and Assessment.}
The comparator episode involves a similar acquisition problem: planning for long-term sustainment and iterative updates of the flight-planning capability. In that case, lifecycle management primarily concerns software maintenance, backlog refinement, and interface stability for deterministic functionality. While this includes managing external data providers, the focus remains on ensuring API uptime and schema consistency rather than monitoring the underlying distribution of the data itself. Integration efforts are centered on building interfaces to known providers, where update cadences are predictable and rarely require full re-validation of the system’s core logic. Consequently, the same general product-support artifacts are invoked, and the SWP-centered governance stack is sufficient to govern these activities.

\paragraph{Comparative Interpretation.}
The comparator shows that SWP is generally adequate when sustainment centers on conventional software maintenance, versioning, and interface stability. In the \emph{Targeting AI} baseline, lifecycle management must also govern monitoring, retraining, and re-authorization as performance conditions evolve, making the absence of clear decision gates and drift-detection triggers harder for program offices to manage consistently.

\subsection{Human Oversight}
\label{subsec:ethical_governance}
\textbf{Finding: }Coverage for human oversight is coded as Partial due to a structural integration gap; while high-level Responsible AI (RAI) guidance is extensive, the pathway fails to operationalize these principles into mandatory Planning Phase artifacts or review gates.
\subsubsection{Baseline Scenario: Targeting AI} 
\paragraph{Circumstance.}
In this episode, the \emph{Targeting AI} program must plan for the human oversight, use constraints, and governance safeguards needed to deploy an AI-enabled targeting capability in an operationally consequential setting. The system’s outputs may shape targeting decisions under time pressure and uncertain conditions. These conditions raise acquisition-time questions about who is accountable for operational use, how intended use and limitations will be defined, what evidence is needed to support responsible deployment, and what mechanisms must exist for human supervision, override, or disengagement when the system behaves unexpectedly.

\paragraph{Policy Actions Invoked.}
The core SWP guidance does not include a dedicated pathway artifact specifically for AI human oversight. This requirement instead comes from responsible-AI guidance within the broader SWP-centered governance stack, particularly the DoD AI Ethical Principles, the 2021 implementation memorandum, and the 2022 Responsible AI (RAI) Strategy and Implementation Pathway \cite{ai_ethical_principles_2020,rai_memo_2021,DoD_RAI_SIP_2022}. Together, these documents provide the principal policy basis for translating high-level commitments into acquisition-relevant governance expectations.

\paragraph{Assessment.}
At the episode-by-concern level, coverage for human oversight is \emph{Partial}. When evaluated against our concern whether policy supports human oversight and responsible or ethical compliance mechanisms appropriate to the use context including accountability, override, and disengagement, the governance stack exhibits a significant translation failure.

The SWP-centered governance stack provides substantial direction through the DoD AI Ethical Principles and the RAI Strategy and Implementation Pathway. These documents establish operational expectations for accountability, use constraints, and human-in-the-loop mechanisms~\cite{ai_ethical_principles_2020,rai_memo_2021,DoD_RAI_SIP_2022}. However, support for human oversight within the core pathway is coded as Partial because the SWP does not translate these high-level commitments into a dedicated artifact requirement or standard decision criterion.

In the Targeting AI case, this omission means that while the Program Manager is subject to RAI policy, there is no fixed mechanism within DoDI 5000.87 to verify that meaningful human oversight is technically enabled before entering the Execution Phase. The principal weakness is structural: the pathway lacks the connective tissue required to move RAI from a statement of intent to an actionable acquisition control.

\paragraph{Gap and Downstream Consequence.}
The gap in this episode is the lack of structural integration of human-oversight expectations into core pathway artifacts and review gates. Because SWP does not mandate an Oversight Plan or specific RAI review gates, program actionability depends entirely on local tailoring and supplemental guidance. The downstream consequence for high-consequence systems like \emph{Targeting AI} is the risk of deploying a capability that lacks documented use limits or explicit technical mechanisms for human override, potentially leading to unintended system consequences in a mission environment.

\subsubsection{Comparator Scenario: Flight Planning Optimizer}
\paragraph{Episode Summary and Assessment.}
In the comparator, the program office must ensure that the flight-planning capability remains subject to operator review. Human review in this case is centered on functional verification: confirming that deterministic software has correctly calculated a route using fixed inputs such as fuel, weight, and weather. Because the logic is transparent and the outputs are reproducible, the core SWP guidance is generally sufficient to support effective human oversight through ordinary usability, verification, and correctness-oriented planning artifacts.

\paragraph{Comparative Interpretation.}
For the flight-planning comparator, operator review is mainly a matter of checking correct execution against transparent logic and known inputs. In \emph{Targeting AI}, operator involvement carries a broader governance function: users must supervise probabilistic recommendations, recognize when the system is operating outside expected conditions, and retain meaningful authority to override or disengage. Those demands are not clearly embedded in standard SWP artifacts.

\section{Discussion}
The discussion proceeds in three parts. In \cref{subsec:actionability_problem}, we identify the core actionability problem in the SWP-centered governance stack. In \cref{subsec:improvements}, we consider what DoD could do about that problem within SWP, including both an AI-supporting sub-path and targeted artifact refinements. Finally, in \cref{subsec:international}, we consider broader implications beyond SWP reform, including institutionalizing lessons learned and drawing comparative insights from external AI procurement frameworks..

\subsection{Actionability Problem in the SWP-Centered Governance Stack}
\label{subsec:actionability_problem}

The episode findings in~\cref{subsec:data_rights,subsec:tevv,subsec:cyber_traceability,subsec:lifecycle_management,subsec:ethical_governance} show that the central problem in the SWP-centered governance stack is not the total absence of AI-relevant policy, but the inconsistent translation of that policy into core pathway guidance. Across the baseline episodes, acquisition teams can often locate relevant guidance somewhere in the broader corpus, yet still lack sufficiently explicit direction on what must be planned, documented, or reviewed at the program level. The result is a recurring actionability problem: AI-relevant controls are often present in principle, but not consistently embedded in the program-facing mechanisms through which SWP is executed. For example, while enterprise-level data strategies exist, they are not translated into specific SWP deliverables for dataset labeling or versioning (\cref{subsec:data_rights}).

This matters because tailoring is not equally sufficient across capability types. In the conventional software comparator, established software artifacts and review practices provided a sufficient governance floor. In the Targeting AI baseline, by contrast, acquisition success depended on AI-specific lifecycle controls that were not yet fully embedded in core SWP artifacts. A concrete example is model drift~(see \cref{subsec:lifecycle_management}): in principle, a program could address it through local tailoring by specifying monitoring metrics, retraining triggers, and rollback criteria. In practice, however, if pathway guidance does not make those expectations visible, the government must already know to ask for them; otherwise, programs are more likely to rely on contractor-defined approaches or local expertise. Because AI practices are still maturing, they necessitate more prescriptive guidance than conventional software to effectively institutionalize emerging capabilities~\cite{brunsson2000world}. Without a solid governance floor, effective controls remain siloed in individual program offices rather than becoming repeatable acquisition norms.

\subsection{How SWP Could Better Support the Acquisition of AI-Enabled Capabilities}
\label{subsec:improvements}

The findings of our Policy Scenario Assessment (\cref{subsec:actionability_problem}) suggest that strengthening SWP support for AI acquisition requires more explicit translation of AI-relevant expectations into pathway structure and program-facing artifacts. Within SWP itself, two responses appear especially important: providing a clearer institutional home for AI-enabled programs and refining the planning artifacts through which AI-relevant controls are implemented. The following subsections consider each response in turn.

\subsubsection{An AI-Supporting Sub-Path Within SWP}
\label{sub-path}

DoDI~5000.87 currently organizes SWP around \emph{applications} and \emph{embedded software} sub-paths, distinguishing software primarily by where and how it is delivered. The actionability problem identified in~\cref{subsec:actionability_problem}, however, suggests that some AI-enabled programs are not fully served by either sub-path as currently structured. Their acquisition burdens arise not only from deployment context, but from the need to plan, document, and sustain AI-specific controls that conventional software programs may not require. One possible institutional response would be an AI-supporting sub-path within SWP that makes these recurring expectations explicit through defined artifacts, content requirements, and review logic. Such a sub-path would provide a clearer structural home for the controls this study found to be underspecified and reduce the burden on program offices to assemble and interpret guidance across multiple documents.

This issue becomes more significant as DoD investment in AI-enabled capabilities grows. As discussed in \cref{SWvsAI}, recent budget materials show a substantial increase from earlier AI-specific funding to a much larger autonomy and AI portfolio concentrated in mission areas such as aerial autonomy, maritime systems, and AI-enabled infrastructure. For some of these acquisitions, the kinds of data, assurance, and lifecycle controls examined in this study are likely to be central to acquisition success. A more explicit pathway for AI-enabled programs would help ensure that those expectations are treated as standard acquisition concerns rather than as local tailoring.

\subsubsection{Targeted Artifact Refinements for AI Acquisition}
\label{subsec:artifacts}
If SWP were to establish an AI-supporting sub-path, that sub-path would need to define more explicit expectations for the planning artifacts through which AI-relevant controls are translated into program practice. The baseline findings in~\cref{subsec:data_rights,subsec:lifecycle_management,subsec:cyber_traceability,subsec:ethical_governance} identify several areas where current SWP guidance remains underspecified for AI-enabled systems. The following refinements illustrate the kinds of artifact-level expectations that an AI-supporting sub-path could formalize.

\begin{description}
    \item[AI-Enhanced Data Strategy:] The data-governance findings show that current planning artifacts do not specify provenance, labeling quality, versioning, or sustainment-oriented data deliverables (\cref{subsec:data_rights}). The Data Strategy should therefore move beyond general references to stewardship by requiring an \emph{AI Data Annex} that specifies expectations for training and evaluation datasets.
    
    \item[Model-Based Product Support Strategy:] The lifecycle-management findings show that monitoring, retraining, and re-authorization triggers remain only partially embedded in current sustainment guidance (\cref{subsec:lifecycle_management}). For AI-capability acquisition programs, the Product Support Strategy (PSS) guidance should therefore be updated to require that planning artifacts explicitly define the triggers for monitoring, retraining, rollback, and re-authorization.
    
    \item[Integrated Cybersecurity and Human Oversight:] The findings for cybersecurity and human oversight reveal that AI release provenance and responsible-AI (RAI) expectations are not yet effectively operationalized in standard artifacts or review gates (\cref{subsec:cyber_traceability,subsec:ethical_governance}). For instance, the lack of a mandatory provenance chain linking specific data versions to model weights creates a critical visibility gap~ (\cref{subsec:cyber_traceability}). To address this, these expectations should be codified as mandatory content within the AI sub-path.
\end{description}

These refinements illustrate the kinds of expectations that an AI-supporting sub-path within SWP could formalize. By specifying where AI-relevant controls belong in planning artifacts and how they should be reviewed, such a sub-path would make AI-enabled acquisition more consistent within the existing SWP structure.

\subsection{Implications Beyond SWP Reform}
\label{subsec:international}
The recommendations above focus on reforms that could be implemented within SWP itself. The findings also suggest broader implications for how AI acquisition governance may need to evolve over time. In particular, they raise questions about how DoD institutionalizes lessons from program experience and how developments in external AI procurement frameworks may inform future acquisition practice.

\subsubsection{Institutionalizing Lessons Learned for AI Acquisition}

As discussed in~\cref{subsec:actionability_problem}, local tailoring is not inherently undesirable; it is one way SWP accommodates variation across programs. The concern is that when AI-enabled acquisition depends too heavily on program-level expertise and interpretation, effective solutions may remain siloed within individual offices rather than becoming visible and reusable across the Department.

For that reason, improving SWP for AI-enabled acquisition should involve more than a lessons-learned repository or reporting requirement. Recent evidence from high-reliability software organizations suggests that when lessons learned remain informal, ad hoc, or weakly tied to operational decision points, failures recur and experience remains fragmented~\cite{Anandayuvaraj2026LearningFailures}. A more effective approach would be to institutionalize lessons learned at defined pathway review points so that emerging practices are assessed and incorporated into future acquisition decisions. Such a function could be housed within organizations such as CDAO or DAU to help ensure that program experience informs future acquisition norms rather than remaining isolated within individual efforts.

\subsubsection{Comparative Notes from External AI Procurement Frameworks}
As discussed in \cref{sec:intro}, governments are developing AI procurement guidance that makes expectations for testing, transparency, and lifecycle oversight more explicit~\cite{japan_procurement_guidelines,eu_ai_act}. This trend is reflected in binding regulations, procurement-oriented guidance, and statutory frameworks that regulate vendor obligations, including the EU AI Act and South Korea’s AI statutory frameworks~\cite{eu_ai_act,skorea_ai_act}. For the DoD, these developments offer useful examples of how broad AI-governance principles can be translated into procurement-facing requirements. By observing how other public-sector buyers formalize those expectations, the Department can refine the technical gates and oversight mechanisms within SWP and make AI-enabled acquisition more explicit and repeatable.

These comparative developments may also carry implications beyond the Department, because U.S. defense acquisition approaches often influence allied reform efforts. Recent partner-nation modernization work has used U.S. acquisition practice as a reference point for structuring software and rapid capability acquisition \cite{south_korea_acquisition_reform}. As a result, a clearer SWP framework for AI-enabled capabilities could matter not only for DoD programs, but also as a more legible and actionable model for partner nations seeking to modernize their own defense acquisition processes.

\section{Conclusion}

This paper asked: \emph{To what extent does the Software Acquisition Pathway (SWP), as implemented through the broader SWP-centered governance stack, accommodate AI’s distinctive technical and operational characteristics, and where does it remain underspecified?} Our analysis indicates that the current governance structure provides a workable foundation for acquiring AI-enabled capabilities, but remains uneven in how it translates high-level AI policy into program-facing acquisition practice. The core weakness identified in this study is a policy-to-artifact gap: although AI-relevant controls exist within the broader governance stack, they are not consistently translated into artifacts, content expectations, and review logic needed to govern programs effectively under SWP. This gap can be bridged by establishing a clear institutional home for AI acquisition programs within SWP, together with explicit artifact-level requirements for the AI-relevant controls that acquisition teams must plan, document, and review.

\paragraph{Future Work.}
This study is intentionally bounded to the SWP's Planning Phase governance and two representative scenarios. Future work should examine how these planning-phase gaps manifest during execution, including model updates, release decisions, and operational monitoring under continuous delivery conditions. It should also compare formal pathway expectations with real acquisition programs, contracts, and solicitations to assess how program offices instantiate AI governance in practice.


\begin{acks}
The views expressed in this article are those of the authors and do not reflect the official policy or position of the  United States Space Force, Department of the Air Force, Department of Defense, or the U.S. Government.
\end{acks}

\clearpage
\bibliographystyle{ACM-Reference-Format}
\bibliography{references_not_zotero}

@online{dau_aaf_policies,
  title   = {Acquisition Policies | Adaptive Acquisition Framework},
  author  = {{Defense Acquisition University}},
  url     = {https://aaf.dau.edu/policy/},
  urldate = {2025-11-10},
  year    = {2025}
}

@misc{KarpathySW2,
  author       = {Karpathy, Andrej},
  title        = {Software 2.0},
  year         = {2017},
  month        = nov,
  day          = {11},
  howpublished = {\url{https://karpathy.medium.com/software-2-0-a64152b37c35}}
}

@online{cio_ai_acquisition_policy,
  title   = {Resources},
  author  = {{U.S. Federal Chief Information Officers Council}},
  url     = {https://www.councils.gov/resources/},
  urldate = {2025-11-10},
  year    = {2025}
}

@online{dow_publications_software_pathway,
  title   = {Department of Defense — Publications (Search: “Software Pathway”, since Jan 2021)},
  author  = {{U.S. Department of Defense}},
  url     = {https://www.war.gov/News/Publications/StartDate/2020-01-01/?Search=software+pathway},
  urldate = {2025-11-10},
  year    = {2025}
}

@online{GAO-23-105850,
  title   = {Artificial Intelligence: DoD Needs Department-Wide Guidance to Inform Acquisitions},
  author  = {{U.S. Government Accountability Office}},
  year    = {2023},
  url     = {https://www.gao.gov/products/gao-23-105850},
  urldate = {2025-11-10}
}

@techreport{DoDI5000_02,
  author       = {Office of the Under Secretary of Defense for Acquisition and Sustainment},
  title        = {Operation of the Adaptive Acquisition Framework},
  institution  = {U.S. Department of Defense},
  type         = {DoD Instruction 5000.02},
  year         = {2020},
  month        = jan,
  day          = {23},
  url          = {https://www.esd.whs.mil/Portals/54/Documents/DD/issuances/dodi/500002p.pdf?ver=2020-01-23-144114-093}
}

@techreport{DoDI5000_87,
  author       = {Office of the Under Secretary of Defense for Acquisition and Sustainment},
  title        = {Operation of the Software Acquisition Pathway},
  institution  = {U.S. Department of Defense},
  type         = {DoD Instruction 5000.87},
  year         = {2020},
  month        = oct,
  day          = {2},
  url          = {https://www.esd.whs.mil/Portals/54/Documents/DD/issuances/dodi/500087p.pdf}
}

@techreport{DoDI5000_88,
  author       = {Office of the Under Secretary of Defense for Research and Engineering},
  title        = {Engineering of Defense Systems},
  institution  = {U.S. Department of Defense},
  type         = {DoD Instruction 5000.88},
  year         = {2020},
  month        = nov,
  day          = {18},
  url          = {https://www.esd.whs.mil/Portals/54/Documents/DD/issuances/dodi/500088p.PDF}
}

@techreport{DoDI5000_89,
  author       = {U.S. Department of Defense},
  title        = {DoD Instruction 5000.89: Test and Evaluation},
  institution  = {Office of the Under Secretary of Defense for Research and Engineering and Office of the Director, Operational Test and Evaluation},
  year         = {2020},
  month        = nov,
  day          = {19},
  url          = {https://www.esd.whs.mil/Portals/54/Documents/DD/issuances/dodi/500089p.PDF}
}

@techreport{LaPlante2022SWPDBS,
  author       = {LaPlante, William A.},
  title        = {Use of the Software Acquisition Pathway for Defense Business Systems},
  institution  = {Office of the Under Secretary of Defense for Acquisition and Sustainment},
  type         = {Memorandum},
  number       = {USA000671-22},
  address      = {3010 Defense Pentagon, Washington, DC 20301-3010},
  year         = {2022},
  month        = aug,
  day          = {24},
  url          = {https://aaf.dau.edu/storage/2022/09/USDA-S-signed-Memo-Use-of-the-SWP-for-DBS_20220824.pdf}
}

@techreport{DoD2025SoftwareAcquisition,
  author       = {Secretary of Defense},
  title        = {Directing Modern Software Acquisition to Maximize Lethality},
  institution  = {U.S. Department of Defense},
  address      = {Washington, DC},
  year         = {2025},
  month        = mar,
  day          = {6},
  type         = {Memorandum},
  url          = {https://media.defense.gov/2025/Mar/07/2003662943/-1/-1/1/DIRECTING-MODERN-SOFTWARE-ACQUISITION-TO-MAXIMIZE-LETHALITY.PDF}
}

@techreport{csis2023pathforward,
  title        = {A New Path Forward: An Analysis of Current AI Software Acquisition Procedures},
  author       = {Sayler, Kelley and Hunter, Andrew and Scharre, Paul},
  year         = {2023},
  institution  = {Center for Strategic and International Studies (CSIS)},
  address      = {Washington, D.C.},
  url          = {https://www.csis.org/analysis/new-path-forward-ai-software-acquisition}
  
}

@inproceedings{baylor_tfx_2017,
  author = {Baylor, Denis and Breck, Eric and Cheng, Heng-Tze and Fiedel, Noah and Foo, Chuan Yu and Haque, Zakaria and Haykal, Salem and Ispir, Mustafa and Jain, Vihan and Koc, Levent and Koo, Chiu Yuen and Lew, Lukasz and Mewald, Clemens and Modi, Akshay Naresh and Polyzotis, Neoklis and Ramesh, Sukriti and Roy, Sudip and Whang, Steven Euijong and Wicke, Martin and Wilkiewicz, Jarek and Zhang, Xin and Zinkevich, Martin},
  title = {TFX: A TensorFlow-Based Production-Scale Machine Learning Platform},
  booktitle = {Proceedings of the 23rd ACM SIGKDD International Conference on Knowledge Discovery and Data Mining},
  year = {2017},
  isbn = {9781450348874},
  pages = {1387–1395},
  numpages = {9},
  doi = {10.1145/3097983.3098021}
}

@inproceedings{MLTestScore,
  author    = {Breck, Eric and Cai, Shanqing and Nielsen, Eric and Salib, Michael and Sculley, D.},
  title     = {The ML Test Score: A Rubric for ML Production Readiness and Technical Debt Reduction},
  booktitle = {2017 IEEE International Conference on Big Data (Big Data)},
  year      = {2017},
  pages     = {1123--1132},
  doi       = {10.1109/BigData.2017.8258038},
  url       = {https://research.google.com/pubs/archive/aad9f93b86b7addfea4c419b9100c6cdd26cacea.pdf}
}

@article{sculley_hidden_debt_2015,
  author  = {Sculley, D. and Holt, G. and Golovin, D. and Davydov, E. and Phillips, T. and Ebner, D. and Chaudhary, V. and Young, M. and Crespo, J.-F. and Dennison, D.},
  title   = {Hidden Technical Debt in Machine Learning Systems},
  journal = {Advances in Neural Information Processing Systems},
  year    = {2015},
  volume  = {28}
}

@article{gebru_datasheets_2021,
  author  = {Gebru, Timnit and Morgenstern, Jamie and Vecchione, Briana and Vaughan, Jennifer Wortman and Wallach, Hanna and Daum{\'e} III, Hal and Crawford, Kate},
  title   = {Datasheets for Datasets},
  journal = {Communications of the ACM},
  year    = {2021},
  volume  = {64},
  number  = {12},
  pages   = {86--92},
  doi     = {10.1145/3458723}
}

@inproceedings{mitchell_model_cards_2019,
  author    = {Mitchell, Margaret and Wu, Simone and Zaldivar, Andrew and Barnes, Parker and Vasserman, Lucy and Hutchinson, Ben and Spitzer, Elena and Raji, Inioluwa Deborah and Gebru, Timnit},
  title     = {Model Cards for Model Reporting},
  booktitle = {FAT* '19: Conference on Fairness, Accountability, and Transparency},
  year      = {2019},
  pages     = {220--229},
  doi       = {10.1145/3287560.3287596}
}

@misc{aiindex2025,
  title   = {Artificial Intelligence Index Report 2025},
  author  = {Maslej, Neal and others},
  institution = {Stanford Institute for Human-Centered Artificial Intelligence},
  year    = {2025},
  url     = {https://hai.stanford.edu/assets/files/hai_ai_index_report_2025.pdf}
}

@misc{tfx_mlopsguide_2024,
  title  = {MLOps: Continuous Delivery and Automation Pipelines in Machine Learning},
  author = {{Google Cloud Architecture Center}},
  year   = {2024},
  month  = aug,
  howpublished = {\url{https://cloud.google.com/architecture/mlops-continuous-delivery-and-automation-pipelines-in-machine-learning}}
}

@inproceedings{jiang2023pretrainedreuse,
author = {Jiang, Wenxin and Synovic, Nicholas and Hyatt, Matt and Schorlemmer, Taylor R. and Sethi, Rohan and Lu, Yung-Hsiang and Thiruvathukal, George K. and Davis, James C.},
title = {An Empirical Study of Pre-Trained Model Reuse in the Hugging Face Deep Learning Model Registry},
year = {2023},
isbn = {9781665457019},
publisher = {IEEE Press},
url = {https://doi.org/10.1109/ICSE48619.2023.00206},
doi = {10.1109/ICSE48619.2023.00206},
booktitle = {Proceedings of the 45th International Conference on Software Engineering},
pages = {2463–2475},
numpages = {13},
location = {Melbourne, Victoria, Australia},
series = {ICSE '23}
}

@inproceedings{intoto2019,
  author    = {Santiago Torres-Arias and Hammad Afzali and Trishank Karthik Kuppusamy and Reza Curtmola and Justin Cappos},
  title     = {in-toto: Providing farm-to-table guarantees for bits and bytes},
  booktitle = {28th {USENIX} Security Symposium ({USENIX} Security 19)},
  year      = {2019},
  address   = {Santa Clara, CA},
  pages     = {1393--1410},
  publisher = {{USENIX} Association},
  month     = aug,
  isbn      = {978-1-939133-06-9},
  url       = {https://www.usenix.org/system/files/sec19-torres-arias.pdf}
}

@inproceedings{Purvish_24,
author = {Jajal, Purvish and Jiang, Wenxin and Tewari, Arav and Kocinare, Erik and Woo, Joseph and Sarraf, Anusha and Lu, Yung-Hsiang and Thiruvathukal, George K. and Davis, James C.},
title = {Interoperability in Deep Learning: A User Survey and Failure Analysis of ONNX Model Converters},
year = {2024},
isbn = {9798400706127},
publisher = {Association for Computing Machinery},
address = {New York, NY, USA},
url = {https://doi.org/10.1145/3650212.3680374},
doi = {10.1145/3650212.3680374},
booktitle = {Proceedings of the 33rd ACM SIGSOFT International Symposium on Software Testing and Analysis},
pages = {1466–1478},
numpages = {13},
location = {Vienna, Austria},
series = {ISSTA 2024}
}

@misc{slsa_v1_2023,
  author = {{SLSA Community}},
  title  = {Supply-chain Levels for Software Artifacts (SLSA) v1.0},
  year   = {2023},
  url    = {https://slsa.dev/spec/v1.0}
}

@misc{sigstore2021,
  author = {{Sigstore}},
  title  = {sigstore: Software Signing for Everyone},
  year   = {2021},
  url    = {https://www.sigstore.dev}
}

@article{provenance_XAI,
  author = {Zhang, Jiachi and Zhou, Wenchao and Ujcich, Benjamin E.},
  title = {Provenance-Enabled Explainable AI},
  year = {2024},
  issue_date = {December 2024},
  publisher = {Association for Computing Machinery},
  address = {New York, NY, USA},
  volume = {2},
  number = {6},
  url = {https://doi.org/10.1145/3698826},
  doi = {10.1145/3698826},
  journal = {Proc. ACM Manag. Data},
  month = dec,
  articleno = {250},
  numpages = {27}
}

@inproceedings{scoresheet_XAI,
  author = {Winikoff, Michael and Thangarajah, John and Rodriguez, Sebastian},
  title = {A Scoresheet for Explainable AI},
  year = {2025},
  isbn = {9798400714269},
  publisher = {International Foundation for Autonomous Agents and Multiagent Systems},
  address = {Richland, SC},
  booktitle = {Proceedings of the 24th International Conference on Autonomous Agents and Multiagent Systems},
  pages = {2171--2180},
  numpages = {10},
  location = {Detroit, MI, USA},
  series = {AAMAS '25}
}

@article{AIgov,
  author = {Mellouli, Sehl and Janssen, Marijn and Ojo, Adegboyega},
  title = {Introduction to the Issue on Artificial Intelligence in the Public Sector: Risks and Benefits of AI for Governments},
  year = {2024},
  issue_date = {March 2024},
  publisher = {Association for Computing Machinery},
  address = {New York, NY, USA},
  volume = {5},
  number = {1},
  url = {https://doi.org/10.1145/3636550},
  doi = {10.1145/3636550},
  journal = {Digit. Gov.: Res. Pract.},
  month = mar,
  articleno = {1},
  numpages = {6}
}

@article{2023ContinuousAI,
  author    = {Steidl, Monika and Felderer, Michael and Ramler, Rudolf},
  title     = {The Pipeline for the Continuous Development of Artificial Intelligence Models: Current State of Research and Practice},
  journal   = {Journal of Systems and Software},
  year      = {2023},
  volume    = {199},
  pages     = {111615},
  doi       = {10.1016/j.jss.2023.111615},
  url       = {https://arxiv.org/abs/2301.09001}
}

@misc{japan_procurement_guidelines,
  title  = {Guidelines for Japanese Government's Procurement and Utilization of Generative AI (Provisional Translation)},
  author = {{Digital Agency of Japan}},
  year   = {2025},
  month  = may,
  url    = {https://www.digital.go.jp/assets/contents/node/basic_page/field_ref_resources/e2a06143-ed29-4f1d-9c31-0f06fca67afc/6e45a64f/20250527_resources_standard_guidelines_guideline_04.pdf}
}

@inproceedings{health,
  author    = {Roy, Soumyadeep and Sundaram, Sowmya S. and Wolff, Dominik and Ganguly, Niloy},
  title     = {Building Trustworthy AI Models for Medicine: From Theory to Applications},
  booktitle = {Proceedings of the Eighteenth ACM International Conference on Web Search and Data Mining},
  series    = {WSDM '25},
  year      = {2025},
  address   = {Hannover, Germany},
  pages     = {1012--1015},
  publisher = {Association for Computing Machinery},
  doi       = {10.1145/3701551.3703477},
  url       = {https://doi.org/10.1145/3701551.3703477}
}

@inproceedings{education,
  author    = {Velazquez-Garcia, Lydia and Cedillo-Hernandez, Antonio and Longar-Blanco, Maria Del Pilar and Bustos-Farias, Eduardo},
  title     = {Enhancing Educational Gamification through AI in Higher Education},
  booktitle = {Proceedings of the 2024 16th International Conference on Education Technology and Computers},
  series    = {ICETC '24},
  year      = {2025},
  publisher = {Association for Computing Machinery},
  address   = {New York, NY, USA},
  pages     = {213--218},
  numpages  = {6},
  isbn      = {9798400717819},
  doi       = {10.1145/3702163.3702416},
  url       = {https://doi.org/10.1145/3702163.3702416}
}

@techreport{dau2022aiacquisition,
  author      = {Pellerin, Cheryl and Wood, Stephen and Allen, Mark},
  title       = {Artificial Intelligence (AI) and Machine Learning (ML) Acquisition and Policy Implications},
  institution = {Defense Acquisition University (DAU)},
  address     = {Fort Belvoir, VA},
  year        = {2022},
  url         = {https://www.dau.edu/library/arj/p/AI-ML-Acquisition-Implications}
}

@techreport{SWAP2019,
  author       = {Defense Innovation Board},
  title        = {Software Is Never Done: Refactoring the Acquisition Code for Competitive Advantage},
  institution  = {U.S. Department of Defense},
  year         = {2019},
  month        = apr,
  url          = {https://media.defense.gov/2019/Apr/30/2002124828/-1/-1/0/SOFTWAREISNEVERDONE_REFACTORINGTHEACQUISITIONCODEFORCOMPETITIVEADVANTAGE_FINAL.SWAP.REPORT.PDF}
}

@misc{DoDI5000_83,
  author       = {{U.S. Department of Defense}},
  title        = {{DoD Instruction 5000.83: Technology and Program Protection to Maintain Technological Advantage}},
  year         = {2020},
  month        = jul,
  howpublished = {\url{https://www.esd.whs.mil/Portals/54/Documents/DD/issuances/dodi/500083p.pdf}}
}

@misc{AIAdoption2023,
  author       = {{Chief Digital and Artificial Intelligence Office}},
  title        = {{2023 Data, Analytics, and Artificial Intelligence Adoption Strategy}},
  year         = {2023},
  month        = nov,
  howpublished = {\url{https://media.defense.gov/2023/Nov/02/2003333300/-1/-1/1/dod_data_analytics_ai_adoption_strategy.pdf}}
}

@misc{DoDI5000_82,
  author       = {U.S. Department of Defense},
  title        = {DoD Instruction 5000.82: Requirements for the Acquisition of Digital Capabilities},
  year         = {2023},
  month        = jun,
  howpublished = {\url{https://www.esd.whs.mil/Portals/54/Documents/DD/issuances/dodi/500082p.pdf}}
}

@inproceedings{jouppi2017tpu,
  author    = {Jouppi, Norman P. and Young, Cliff and Patil, Nishant and Patterson, David and Agrawal, Gaurav and Bajwa, Raminder and Bates, Sarah and Bhatia, Suresh and Boden, Nan and Borchers, Al and Boyle, Rick et al.},
  title     = {In-Datacenter Performance Analysis of a Tensor Processing Unit},
  booktitle = {Proceedings of the 44th Annual International Symposium on Computer Architecture (ISCA '17)},
  year      = {2017},
  pages     = {1--12},
  publisher = {ACM/IEEE},
  doi       = {10.1145/3079856.3080246}
}

@article{DLAccelSurvey2025,
author = {Silvano, Cristina and Ielmini, Daniele and Ferrandi, Fabrizio and Fiorin, Leandro and Curzel, Serena and Benini, Luca and Conti, Francesco and Garofalo, Angelo and Zambelli, Cristian and Calore, Enrico and Schifano, Sebastiano and Palesi, Maurizio and Ascia, Giuseppe and Patti, Davide and Petra, Nicola and De Caro, Davide and Lavagno, Luciano and Urso, Teodoro and Cardellini, Valeria and Cardarilli, Gian Carlo and Birke, Robert and Perri, Stefania},
title = {A Survey on Deep Learning Hardware Accelerators for Heterogeneous HPC Platforms},
year = {2025},
issue_date = {November 2025},
publisher = {Association for Computing Machinery},
address = {New York, NY, USA},
volume = {57},
number = {11},
issn = {0360-0300},
url = {https://doi.org/10.1145/3729215},
doi = {10.1145/3729215},
journal = {ACM Comput. Surv.},
month = jun,
articleno = {286},
numpages = {39}
}

@misc{google2025genaimil,
  title        = {Chief Digital and Artificial Intelligence Office Selects Google Cloud's AI to Power GenAI.mil},
  author       = {{Google Cloud}},
  year         = {2025},
  month        = dec,
  day          = {9},
  howpublished = {\url{https://www.googlecloudpresscorner.com/2025-12-09-Chief-Digital-and-Artificial-Intelligence-Office-Selects-Google-Clouds-AI-to-Power-GenAI-mil}}
}

@misc{army2025palantirEA,
  author       = {{U.S. Army Public Affairs}},
  title        = {U.S. Army Awards Enterprise Service Agreement to Enhance Military Readiness and Drive Operational Efficiency},
  year         = {2025},
  month        = jul,
  day          = {31},
  howpublished = {\url{https://www.army.mil/article/287506/u_s_army_awards_enterprise_service_agreement_to_enhance_military_readiness_and_drive_operational_efficiency}}
}

@incollection{dote2018_mps_jmps_af,
  author       = {{Office of the Director, Operational Test and Evaluation}},
  title        = {Mission Planning System (MPS) / Joint Mission Planning System--Air Force (JMPS-AF)},
  booktitle    = {DOT\&E {FY} 2018 Annual Report},
  institution  = {Office of the Secretary of Defense},
  publisher    = {U.S. Department of Defense},
  year         = {2018},
  pages        = {197--198},
  url          = {https://www.dote.osd.mil/Portals/97/pub/reports/FY2018/af/2018afmps.pdf?ver=2019-08-21-155843-447},
  urldate      = {2026-02-04}
}

@misc{dau_swp_develop_strategies,
  author       = {{Defense Acquisition University}},
  title        = {{Develop Strategies} (Software Acquisition Pathway)},
  organization = {{Defense Acquisition University (DAU)}},
  year         = {n.d.},
  url          = {https://aaf.dau.edu/aaf/software/develop-strategies/},
  urldate      = {2026-02-05}
}

@techreport{DoDI5010_44,
  author       = {{U.S. Department of Defense}},
  title        = {{DoD Instruction 5010.44}: Intellectual Property (IP) Acquisition and Licensing},
  institution  = {Office of the Under Secretary of Defense for Acquisition and Sustainment},
  year         = {2019},
  month        = oct,
  day          = {16},
  url          = {https://www.esd.whs.mil/Portals/54/Documents/DD/issuances/dodi/501044p.pdf},
  urldate      = {2026-02-06}
}

@misc{dau_SW_acq_strategy_2022,
  title        = {Software Acquisition Strategy: Agile Guidance},
  author       = {{Defense Acquisition University (DAU)}},
  organization = {{Defense Acquisition University}},
  year         = {2022},
  month        = jun,
  howpublished = {\url{https://aaf.dau.edu/storage/2022/06/Acquisition-Strategy-Guidance.pdf}}
}

@misc{dafi63-101_2024,
  title        = {Department of the Air Force Instruction 63-101/20-101, \emph{Integrated Life Cycle Management}},
  author       = {{Department of the Air Force}},
  organization = {{Department of the Air Force}},
  year         = {2024},
  month        = feb,
  howpublished = {\url{https://static.e-publishing.af.mil/production/1/saf_aq/publication/dafi63-101_20-101/dafi63-101_20-101.pdf}}
}

@techreport{omb_m25_22_2025,
  author       = {Vought, Russell T.},
  title        = {Driving Efficient Acquisition of Artificial Intelligence in Government},
  institution  = {Office of Management and Budget, Executive Office of the President},
  type         = {Memorandum},
  number       = {M-25-22},
  year         = {2025},
  month        = apr,
  day          = {3},
  url          = {https://www.whitehouse.gov/wp-content/uploads/2025/02/M-25-22-Driving-Efficient-Acquisition-of-Artificial-Intelligence-in-Government.pdf}
}

@techreport{DoDI5000_02_2008,
  author      = {{U.S. Department of Defense}},
  title       = {{DoD Instruction 5000.02: Operation of the Defense Acquisition System}},
  institution = {U.S. Department of Defense},
  type        = {Department of Defense Instruction},
  number      = {5000.02},
  year        = {2008},
  month       = dec,
  day         = {8},
  url         = {https://www.dami.army.pentagon.mil/site/artpc/docs/DoDI%205000_02p.pdf}
}

@techreport{DoDI5000_02_2015,
  author      = {{U.S. Department of Defense}},
  title       = {{DoD Instruction 5000.02: Operation of the Defense Acquisition System}},
  institution = {U.S. Department of Defense},
  type        = {Department of Defense Instruction},
  number      = {5000.02},
  year        = {2015},
  month       = jan,
  day         = {7}
}

@techreport{CDAO_SI_TE_2024,
  author       = {{Chief Digital and Artificial Intelligence Office (CDAO)}},
  title        = {Systems Integration Test and Evaluation of Artificial Intelligence-Enabled Capabilities: What to Consider in a Test \& Evaluation Strategy},
  institution  = {U.S. Department of Defense},
  type         = {CDAO Test and Evaluation Strategy Framework (Guidance and Best Practices)},
  year         = {2024},
  month        = apr,
  url          = {https://www.ai.mil/Portals/137/Documents/Resources%20Page/CDAO_TE_Framework_SI_TES_RELEASED_APRIL_2024-compressed.pdf}
}

@techreport{DoDI5000_90,
  author       = {Office of the Under Secretary of Defense for Acquisition and Sustainment},
  title        = {Cybersecurity for Acquisition Decision Authorities and Program Managers},
  institution  = {U.S. Department of Defense},
  type         = {DoD Instruction 5000.90},
  year         = {2020},
  month        = dec,
  day          = {31},
  url          = {https://www.esd.whs.mil/Portals/54/Documents/DD/issuances/dodi/500090p.PDF}
}

@misc{ai_ethical_principles_2020,
  author       = {{U.S. Department of Defense}},
  title        = {{DoD Adopts Ethical Principles for Artificial Intelligence}},
  year         = {2020},
  month        = feb,
  day          = {24},
  howpublished = {Defense.gov Release},
  url          = {https://www.defense.gov/Newsroom/Releases/Release/Article/2091996/dod-adopts-ethical-principles-for-artificial-intelligence/},
  urldate      = {2026-02-15}
}

@misc{rai_memo_2021,
  author       = {{Deputy Secretary of Defense}},
  title        = {{Implementing Responsible Artificial Intelligence in the Department of Defense}},
  year         = {2021},
  month        = may,
  day          = {26},
  howpublished = {Memorandum},
  url          = {https://media.defense.gov/2021/May/27/2002730593/-1/-1/0/IMPLEMENTING-RESPONSIBLE-ARTIFICIAL-INTELLIGENCE-IN-THE-DEPARTMENT-OF-DEFENSE.PDF},
  urldate      = {2026-02-15}
}

@techreport{cunningham2016_psa,
  author      = {Cunningham, Stephen},
  title       = {Policy Scenario Analysis: A Methodological Overview},
  institution = {Bay of Bengal Programme Inter-Governmental Organisation},
  year        = {2016},
  month       = nov,
  number      = {BOBP/WB/OPP/REP 20},
  address     = {Chennai, Tamil Nadu, India}
}

@inproceedings{looker2008scenario,
  author    = {Looker, Nik and Webster, David and Russell, Duncan and Xu, Jie},
  title     = {Scenario Based Evaluation},
  booktitle = {2008 11th IEEE International Symposium on Object and Component-Oriented Real-Time Distributed Computing (ISORC)},
  year      = {2008},
  pages     = {148--154},
  publisher = {IEEE},
  doi       = {10.1109/ISORC.2008.56}
}

@misc{ProjectMavenMemo2017,
  author = {{U.S. Department of Defense}},
  title  = {{Establishment of an Algorithmic Warfare Cross-Functional Team (Project Maven)}},
  year   = {2017},
  month  = apr,
  url    = {https://dodcio.defense.gov/Portals/0/Documents/Project\%20Maven\%20DSD\%20Memo\%2020170425.pdf}
}

@misc{DoDFY2026SOCOM,
  author       = {{U.S. Department of Defense, United States Special Operations Command}},
  title        = {{Fiscal Year 2026 Budget Estimates: Research, Development, Test \& Evaluation, Defense-Wide, United States Special Operations Command}},
  year         = {2025},
  month        = jun,
  howpublished = {\url{https://comptroller.defense.gov/Portals/45/Documents/defbudget/FY2026/budget_justification/pdfs/03_RDT_and_E/RDTE_SOCOM_PB_2026.pdf}}
}

@misc{xtechOverwatch2025,
  author       = {{U.S. Army xTech Program}},
  title        = {{xTechOverwatch Informational Session Presentation Deck}},
  year         = {2025},
  month        = apr,
  howpublished = {\url{https://xtech.army.mil/wp-content/uploads/2025/04/xTechOverwatch-Informational-Session-Presentation-Deck.pdf}}
}

@misc{PMISAWeb,
  author       = {{Program Executive Office Intelligence, Electronic Warfare \& Sensors}},
  title        = {{Project Manager Intelligence Systems \& Analytics}},
  howpublished = {\url{https://cpeisw.army.mil/pm-isa/}},
  year         = {2026}
}

@misc{DoDC3Strategy2020,
  author       = {{U.S. Department of Defense}},
  title        = {{DoD C3 Modernization Strategy}},
  year         = {2020},
  howpublished = {\url{https://dodcio.defense.gov/Portals/0/Documents/DoD-C3-Strategy.pdf}}
}

@misc{armyPEODemand2025,
  author       = {{U.S. Army xTech Program}},
  title        = {{PEO Demand Signal Slides -- Master Deck}},
  year         = {2025},
  month        = mar,
  howpublished = {\url{https://xtech.army.mil/wp-content/uploads/2025/03/PEO-Demand-Signal-Slides-Master-Deck.pdf}}
}

@inproceedings{Anandayuvaraj2026LearningFailures,
  author    = {Dharun Anandayuvaraj and Tanmay Singla and Zain A. H. Hammadeh and Andreas Lund and Alexandra Holloway and James C. Davis},
  title     = {Learning From Software Failures: A Case Study at a National Space Research Center},
  booktitle = {Proceedings of the 2026 IEEE/ACM 48th International Conference on Software Engineering (ICSE '26)},
  year      = {2026},
  address   = {Rio de Janeiro, Brazil},
  publisher = {ACM},
  doi       = {10.1145/3744916.3773149}
}

@article{Kattel2022,
  author = {Kattel, Rainer and Drechsler, Wolfgang},
  title = {Public Procurement and Innovation: Theory and Practice},
  journal = {Standardization and Public Policy Review},
  volume = {14},
  number = {2},
  pages = {125--143},
  year = {2022},
  publisher = {Springer},
  doi = {10.1007/978-3-642-40258-6}
}

@misc{DAU_AAF_2026,
  author = {{Defense Acquisition University}},
  title = {Adaptive Acquisition Framework (AAF) Overview},
  year = {2026},
  urldate = {2026-02-07},
  url = {https://aaf.dau.edu/}  
}

@misc{DAU_AAF_Software,
  author       = {{Defense Acquisition University}},
  title        = {{Software Acquisition}},
  howpublished = {\url{https://aaf.dau.edu/aaf/software/}}
}

@techreport{dod_fy26_weapons,
  author       = {{Office of the Under Secretary of Defense (Comptroller)}},
  title        = {{FY 2026 Program Acquisition Costs by Weapon System}},
  institution  = {{U.S. Department of Defense}},
  year         = {2025},
  month        = jul,
  url          = {https://comptroller.war.gov/Portals/45/Documents/defbudget/FY2026/FY2026_Weapons.pdf}
}

@misc{cmmi2024_levels,
  author       = {{ISACA}},
  title        = {{CMMI Levels of Capability and Performance}},
  year         = {2024},
  howpublished = {\url{https://cmmiinstitute.com/learning/appraisals/levels}}
}

@book{ragin2008redesigning,
  author    = {Charles C. Ragin},
  title     = {Redesigning Social Inquiry: Fuzzy Sets and Beyond},
  publisher = {University of Chicago Press},
  address   = {Chicago},
  year      = {2008},
  isbn      = {9780226702759},
  doi       = {10.7208/chicago/9780226702797.001.0001}
}

@TECHREPORT{south_korea_acquisition_reform,
title = {Closing the Gap: Modernizing South Korea's Defense Acquisition Framework},
author = {Jang, Won-Joon and Park, Hea Ji},
year = {2024},
institution = {Korea Institute for Industrial Economics and Trade},
type = {Research Papers},
number = {24/6},
url = {https://EconPapers.repec.org/RePEc:ris:kietrp:2024_006}
}

@book{miles_2014,
  author    = {Matthew B. Miles and A. Michael Huberman and Johnny Salda{\~n}a},
  title     = {Qualitative Data Analysis: A Methods Sourcebook},
  edition   = {3rd},
  publisher = {SAGE Publications},
  address   = {Thousand Oaks, CA},
  year      = {2014}
}

@techreport{DoD_RAI_SIP_2022,
  author      = {{U.S. Department of Defense}},
  title       = {Responsible Artificial Intelligence Strategy and Implementation Pathway},
  institution = {Chief Digital and Artificial Intelligence Office (CDAO)},
  address     = {Washington, DC},
  year        = {2022},
  month       = {June},
  type        = {Strategy and Implementation Pathway},
  number      = {AD1215042},
  url         = {https://media.defense.gov/2022/Jun/22/2003022604/-1/-1/0/Department-of-Defense-Responsible-Artificial-Intelligence-Strategy-and-Implementation-Pathway.PDF}
}

@techreport{microsoft_2023_ai_procurement,
  author       = {{Microsoft Worldwide Public Sector}},
  title        = {Advancing AI Procurement and Adoption in the Public Sector: Considerations, Use Cases and Practical Approaches},
  institution  = {Microsoft Worldwide Public Sector},
  year         = {2023},
  month        = dec,
  address      = {Redmond, WA},
  url          = {https://wwps.microsoft.com/wp-content/uploads/2023/12/Microsoft-AI-Procurement-Paper_Final.pdf}
}

@techreport{nist_ai_rmf_1_0,
  author       = {{National Institute of Standards and Technology}},
  title        = {AI Risk Management Framework (AI RMF 1.0)},
  institution  = {National Institute of Standards and Technology},
  number       = {NIST AI 100-1},
  year         = {2023},
  url          = {https://nvlpubs.nist.gov/nistpubs/ai/NIST.AI.100-1.pdf}
}

@techreport{dod_fy25_ai,
  author       = {{Office of the Under Secretary of Defense (Comptroller)}},
  title        = {Defense Budget Overview, Fiscal Year 2025},
  institution  = {Office of the Under Secretary of Defense (Comptroller)},
  year         = {2024},
  address      = {Washington, DC},
  url          = {https://comptroller.defense.gov/Budget-Materials/}
}

@article{eu_ai_act,
  author       = {{European Parliament and Council of the European Union}},
  title        = {Regulation (EU) 2024/1689 of the European Parliament and of the Council of 13 June 2024 laying down harmonised rules on artificial intelligence (Artificial Intelligence Act)},
  journal      = {Official Journal of the European Union},
  year         = {2024},
  month        = jul,
  url          = {https://eur-lex.europa.eu/eli/reg/2024/1689/oj/eng}
}

@techreport{dod_fy26_ai,
  author       = {{Office of the Under Secretary of Defense (Comptroller)}},
  title        = {Defense Budget Overview, Fiscal Year 2026},
  institution  = {Office of the Under Secretary of Defense (Comptroller)},
  year         = {2025},
  address      = {Washington, DC},
  url          = {https://comptroller.defense.gov/Budget-Materials/}
}

@techreport{skorea_ai_act,
  author       = {{Ministry of Government Legislation, Republic of Korea}},
  title        = {Framework Act on the Development of Artificial Intelligence},
  institution  = {Ministry of Government Legislation, Republic of Korea},
  year         = {2025},
  month        = jul,
  url          = {https://cset.georgetown.edu/publication/south-korea-ai-law-2025/}
}

@techreport{gao_software_reform_2021,
  author      = {{U.S. Government Accountability Office}},
  title       = {{DOD Software Acquisition: Status of and Challenges Related to Reform Efforts}},
  institution = {{U.S. Government Accountability Office}},
  number      = {GAO-21-105298},
  address     = {Washington, DC},
  year        = {2021},
  month       = sep,
  url         = {https://www.gao.gov/products/gao-21-105298}
}

@inproceedings{dunlap_good_bad_ugly_2024,
  author    = {Dunlap, Jeffrey},
  title     = {Innovation in Software Acquisition: The Good, Bad, and Ugly},
  booktitle = {Proceedings of the Twenty-First Annual Acquisition Research Symposium},
  institution = {Acquisition Research Program, Naval Postgraduate School},
  address   = {Monterey, CA},
  year      = {2024},
  month     = may,
  url       = {https://dair.nps.edu/bitstream/123456789/5136/1/SYM-AM-24-073.pdf}
}

@techreport{tate_bailey_feasible_2022,
  author      = {Tate, David M. and Bailey, John},
  title       = {When is it Feasible (or Desirable) to Use the Software Acquisition Pathway?},
  institution = {Acquisition Research Program, Naval Postgraduate School},
  address     = {Monterey, CA},
  year        = {2022},
  month       = may,
  url         = {https://www.dair.nps.edu/handle/123456789/4569}
}

@book{brunsson2000world,
  author    = {Brunsson, Nils and Jacobsson, Bengt},
  title     = {A World of Standards},
  publisher = {Oxford University Press},
  year      = {2000},
  isbn      = {978-0198297598}
}

\appendix

\end{document}